\documentclass[preprint,showpacs,preprintnumbers,amsmath,amssymb]{revtex4}

\usepackage{graphicx}
\usepackage{dcolumn}
\usepackage{bm}

\begin{document}

\preprint{}

\title{Qubit-Qubit and Qubit-Qutrit Separability Functions and Probabilities}

\author{Paul B. Slater}%
\email{slater@kitp.ucsb.edu}
\affiliation{%
ISBER, University of California, Santa Barbara, CA 93106\\
}%
\date{\today}

\begin{abstract}
We list in increasing order --- $\left\{\frac{1}{10}, 
\frac{1}{3},\frac{3}{8},\frac{2}{5},\frac{135 \pi
   }{1024},\frac{16}{3 \pi ^2},\frac{3 \pi
   }{16},\frac{5}{8},\frac{105 \pi }{512},2-\frac{435 \pi
   }{1024},\frac{11}{16},1\right\}$ --- a number of exact 
two-qubit {\it Hilbert-Schmidt} (HS) separability probabilities, we are 
able to compute.
Each probability corresponds to a specific
scenario --- a class of $4 \times 4$ density matrices ($\rho$) with 
an $m$-subset ($m<6$)
of its six off-diagonal pairs of symmetrically located entries set
to zero. Initially, we consider only scenarios in which the
$(6-m)$ {\it non}-nullified pairs are real, but then permit them to be
complex (as well as quaternionic) in nature. 
The general analytical strategy we implement is based on
the Bloore density matrix 
parameterization ({\it J. Phys. A}, {{\bf{9}}}, 2059 [1976]), 
allowing us to conveniently reduce the dimensionalities of
required integrations. For each scenario, we 
identify a certain univariate 
``separability function'' $\mathcal{S}_{scenario}(\nu)$, 
where $\nu = \frac{\rho_{11} \rho_{44}}{\rho_{22} \rho_{33}}$. 
The integral over $\nu \in [0,\infty]$ of the product of this
function with a scenario-specific (marginal) jacobian function 
$\mathcal{J}_{scenario}(\nu)$ yields
the HS {\it separable} volume ($V^{HS}_{sep}$). The ratio of 
$V^{HS}_{sep}$ to the
HS {\it total} (entangled and non-entangled) 
volume gives us the HS scenario-specific separability {\it probability}.
Among the possible forms that we 
have so far determined for 
$\mathcal{S}_{scenario}(\nu)$ are piecewise combinations of $c$, 
$c \sqrt{\nu}$ and 
$c \nu$ for $\nu \in [0,1]$ and  (in a dual manner) 
$c$, $\frac{c}{\sqrt{\nu}}$ 
and $\frac{c}{\nu}$ for $\nu \in [1,\infty]$.
We also obtain {\it bivariate} separability functions 
$S^{6 \times 6}_{scenario}(\nu_{1},\nu_{2})$ in the qubit-{\it qutrit} case,
involving $6 \times 6$ density matrices, having ratio variables,
$\nu_{1}= \frac{\rho_{11} \rho_{55}}{\rho_{22} \rho_{44}}$ and 
$\nu_{2}= \frac{\rho_{22} \rho_{66}}{\rho_{33} \rho_{55}}$.
Additionally, we investigate parallel two-qutrit and three-qubit
problems. 
Further still, we find some {\it analytic} evidence for the relevance of
{\it beta} functions to the two-qubit separability function 
problem, while we previously 
({\it Phys. Rev. A} [2007]) had
found {\it numerical} evidence of such a nature.
\end{abstract}


\maketitle
\tableofcontents
\section{Introduction}
In the abstract to their recent comprehensive review, the Horodecki family
note that while quantum entanglement is ``usually fragile to environment, 
it is robust against conceptual and mathematical tools, the task of 
which is to decipher its rich structure'' \cite{family}.
In the study below, we certainly 
encounter such robustness and pursue the task of surmounting 
it --- with some successes.

Given a generic class composed of composite 
quantum systems, the question of the 
``relative proportion'' of entangled and non-entangled states 
in that class  was 
apparently first raised by {\. Z}yczkowski, Horodecki, 
Sanpera and Lewenstein (ZHSL) in a much-cited paper \cite{ZHSL}.
They gave  ``three main reasons'' --- ``philosophical'', ``practical'' 
and ``physical'' --- upon which they elaborated, for pursuing the topic.
The present author, motivated by the ZHSL paper,  
has investigated this issue in a number of  settings, using 
various (monotone and non-monotone) measures on  
quantum states, and a variety of
numerical and analytical methods
\cite{slaterC,slaterA,slaterOptics,slaterqip,
pbsJak,slaterPRA,slaterJGP,pbsCanosa} 
(cf. \cite{sbz,sepsize1,sepsize2,sepsize3}).
Though the problems are challenging 
(high-dimensional) in nature, many of the results obtained 
in answer to the ZHSL question in these various contexts 
have been strikingly simple and elegant 
(and/or conjecturally so).
The new results below certainly fit into
such an interesting, appealing  pattern.

Specifically here, we develop an approach 
recently presented in
\cite{slaterPRA2}. This was found to be relatively effective in
studying the question posed by ZHSL, in the context of two-qubit
systems (the smallest possible example exhibiting entanglement), 
endowed with the
(non-monotone \cite{ozawa}) 
{\it Hilbert-Schmidt}  (HS) measure \cite{szHS}, inducing the flat,
{\it Euclidean} geometry on the space of $4 \times 4$ density matrices.
This approach \cite{slaterPRA2} exploits two distinct features of a
form of density matrix parameterization first discussed by Bloore 
\cite{bloore}. These properties  allow 
us to deal with lower-dimensional integrations
(more amenable to computation) than would otherwise be possible.
We further find that the interesting advantages 
of the Bloore parameterization do, in fact,  carry over --- in a 
somewhat modified fashion --- to the qubit-qutrit 
(sec.~\ref{qubitqutrit}), qutrit-qutrit (sec.~\ref{qutritqutrit}) 
and qubit-qubit-qubit (secs.~\ref{qubitqubitqubitI} and 
\ref{qubitqubitqubitII}) domains.
\subsection{Bloore (off-diagonal-scaling) parameterization}
We shall, first, consider the 9-dimensional convex set of 
(two-qubit) $4 \times 4$ density matrices with {\it real} entries,
and parameterize them --- following Bloore \cite{bloore} 
(cf. \cite[p. 235]{kurowicka}) --- as 
\begin{equation} \label{BlooreDenMat}
\rho = \left(
\begin{array}{llll}
 \rho _{11} & z_{12} \sqrt{\rho _{11} \rho _{22}} & z_{13}
   \sqrt{\rho _{11} \rho _{33}} & z_{14} \sqrt{\rho _{11}
   \rho _{44}} \\
 z_{12} \sqrt{\rho _{11} \rho _{22}} & \rho _{22} & z_{23}
   \sqrt{\rho _{22} \rho _{33}} & z_{24} \sqrt{\rho _{22}
   \rho _{44}} \\
 z_{13} \sqrt{\rho _{11} \rho _{33}} & z_{23} \sqrt{\rho
   _{22} \rho _{33}} & \rho _{33} & z_{34} \sqrt{\rho
   _{33} \rho _{44}} \\
 z_{14} \sqrt{\rho _{11} \rho _{44}} & z_{24} \sqrt{\rho
   _{22} \rho _{44}} & z_{34} \sqrt{\rho _{33} \rho _{44}}
   & \rho _{44}
\end{array}
\right).
\end{equation}
One, of course, has the 
standard requirements that $\rho_{ii} \geq 0$ and 
(the unit trace condition) $\Sigma_{i} \rho_{ii} 
= 1$.
Now, three additional necessary conditions
(which can be expressed {\it without} using the diagonal entries,  
due to the  $\rho_{ii} \geq 0$ stipulation) 
that must be fulfilled 
for $\rho$ to be a
density matrix (with all eigenvalues non-negative) are: (1) 
the non-negativity of the determinant (the principal $4 \times 4$ minor),
\begin{equation} \label{firstcondition}
\left(z_{34}^2-1\right) z_{12}^2+2 \left(z_{14}
   \left(z_{24}-z_{23} z_{34}\right)+z_{13}
   \left(z_{23}-z_{24} z_{34}\right)\right)
   z_{12} -z_{23}^2-z_{24}^2-z_{34}^2 +
\end{equation}
\begin{displaymath}
+ z_{14}^2    \left(z_{23}^2-1\right)+z_{13}^2
   \left(z_{24}^2-1\right)+2 z_{23} z_{24} z_{34}+2 z_{13}
   z_{14} \left(z_{34}-z_{23} z_{24}\right)+1 \geq 0;
\end{displaymath}
(2): 
the non-negativity of the leading principal $3 \times 3$ minor,
\begin{equation} \label{secondcondition}
-z_{12}^2+2 z_{13} z_{23} z_{12}-z_{13}^2-z_{23}^2+1 \geq 0;
\end{equation}
and (3): the non-negativity of the principal $2 \times 2$ minors 
(although actually only the $i=j=1$ case is needed, it is natural
to impose them all),
\begin{equation} \label{2by2}
1-z_{ij}^2 \geq 0.
\end{equation}
As noted,
the {\it diagonal} entries of $\rho$ do {\it not} enter into any of 
these constraints --- which taken together are {\it sufficient} to guarantee
the nonnegativity of $\rho$ itself --- as they can be shown to 
contribute only (cancellable) non-negative factors to the
determinant and principal minors. This cancellation property 
is certainly 
a principal virtue of the Bloore parameterization, allowing one to
proceed analytically in lower dimensions than one might 
initially surmise.
(Let us note that, utilizing this parameterization,
we have been able to establish a recent 
conjecture of M{\aa}nsson, Porta Mana and Bj{\"o}rk regarding Bayesian state 
assignment for three-level quantum systems, and, in fact, verify our own
four-level analogue of their conjecture \cite[eq. (52)]{mansson} 
\cite{mansson2}.)

Additionally, implementing the Peres-Horodecki condition
\cite{asher,michal,bruss} requiring
the non-negativity of the {\it partial transposition} of $\rho$,
we have the 
necessary {\it and} sufficient 
condition for the {\it separability} (non-entanglement) of  $\rho$ 
that (4):
\begin{equation} \label{Thirdcontion}
\nu  \left(z_{34}^2-1\right) z_{12}^2+2 \sqrt{\nu }
   \left(\nu  z_{13} z_{14}+z_{23} z_{24}-\sqrt{\nu }
   \left(z_{14} z_{23}+z_{13} z_{24}\right) z_{34}\right)
   z_{12}-z_{23}^2-\nu  z_{34}^2+\nu +
\end{equation}
\begin{displaymath}
 +\nu 
   \left(\left(z_{24}^2-1\right) z_{13}^2-2 z_{14} z_{23}
   z_{24} z_{13}-z_{24}^2+z_{14}^2 \left(z_{23}^2-\nu
   \right)\right)+2 \sqrt{\nu } \left(z_{13} z_{23}+\nu 
   z_{14} z_{24}\right) z_{34} \geq 0,
\end{displaymath}
where
\begin{equation} \label{BlooreRatio}
\nu = \mu^2 = \frac{\rho_{11} \rho_{44}}{\rho_{22} \rho_{33}},
\end{equation}
being the only information needed, at this stage, concerning
the diagonal entries of $\rho$. (It is interesting to contrast
the role of our variable $\nu$, as it pertains to the determination of
entanglement, with the rather different roles played by 
the {\it concurrence} and {\it negativity} \cite{ver2,imai}.)
We have vacillated between the use of $\nu$ and $\mu$ as our principal
variable in our two previous studies \cite{univariate,slaterPRA2}. In 
sec.~\ref{approximate}, we will revert to the use of $\mu$, as it appears that
its use avoids the appearances of square roots which definitely seem
to impede certain Mathematica operations.

So, the Bloore parameterization is 
evidently even further convenient here, in reducing
the apparent dimensionality of the separable volume problem. That is, we 
now have to consider, initially at least, only the separability variable 
$\nu$ rather than {\it three} independent (variable)
diagonal entries. 
(This supplementary feature had not been commented upon 
by Bloore, as he discussed only
$2 \times 2$ and $3 \times 3$ density matrices, and also, obviously,
since the Peres-Horodecki 
separability condition had not yet been formulated.)
The ({\it two} variable --- $\nu_{1}, \nu_{2}$) 
analogue of (\ref{BlooreRatio}) in the $6 \times 6$ 
(qubit-{\it qutrit}) case will be discussed 
and implemented in sec.~\ref{qubitqutrit}.
Additionally still, we find a {\it four}-variable counterpart in
the $9 \times 9$ qutrit-qutrit instance [sec.~\ref{qutritqutrit}], 
and {\it three}-variable counterparts in two sets of qubit-qubit-qubit analyses
(secs.~\ref{qubitqubitqubitI} and \ref{qubitqubitqubitII}).
(The question of whether any or all of these several 
ratio variables are
themselves {\it observables} would appear to be of some interest.)

In \cite[eqs. (3)-(5)]{slaterPRA2}, we expressed the conditions
(found through application of the ``cylindrical algebraic decomposition'' 
\cite{cylindrical})
that --- in terms of the $z_{ij}$'s --- an 
arbitrary  9-dimensional $4 \times 4$ real density matrix 
$\rho$ must fulfill.
These took the form,
\begin{equation} \label{limits}
 z_{12}, z_{13}, z_{14} \in [-1,1],
 z_{23} \in [Z^-_{23},Z^+_{23}],
 z_{24} \in [Z^-_{24},Z^+_{24}],
 z_{34} \in [Z^-_{34},Z^+_{34}],
\end{equation}
where
\begin{equation}
Z^{\pm}_{23} =z_{12} z_{13} \pm \sqrt{1-z_{12}^2} \sqrt{1-z_{13}^2} , 
Z^{\pm}_{24} =z_{12} z_{14} \pm \sqrt{1-z_{12}^2} \sqrt{1-z_{14}^2} ,
\end{equation}
\begin{displaymath}
Z^{\pm}_{34} = \frac{z_{13} z_{14} -z_{12} z_{14} z_{23} -z_{12} z_{13} z_{24} +z_{23}
z_{24} \pm s}{1-z_{12}^2},
\end{displaymath}
and
\begin{equation} \label{limits2}
s = \sqrt{-1 +z_{12}^2 +z_{13}^2 -2 z_{12} z_{13} z_{23} +z_{23}^2}
\sqrt{-1 +z_{12}^2 +z_{14}^2 -2 z_{12} z_{14} z_{24} +z_{24}^2}.
\end{equation}

In his noteworthy paper, Bloore also presented \cite[secs. 6,7]{bloore} 
a quite interesting
discussion of the ``spheroidal'' geometry induced by his parameterization.
This strongly suggests that it might prove useful to reparameterize 
the $z_{ij}$ variables in terms of spheroidal-type coordinates.
Closely following the argument of Bloore --- that is, performing rotations
of the $(z_{13},z_{23})$ and $(z_{14},z_{24})$ 
vectors by $\frac{\pi}{4}$ and recogizing that each pair of so-transformed
variables lay in ellipses with axes of length 
$\sqrt{1 \pm z_{12}}$ --- we were able to 
substantially simply
the forms of these conditions ((\ref{limits})-(\ref{limits2})).

Using the set of transformations (having a jacobian equal to 
$\frac{\left(1-z_{12}^2\right) \gamma _1}{2 \gamma _2}$)
\begin{equation} \label{reparameterization}
z_{13}\to \frac{\left(\sqrt{1-z_{12}} \cos
   \left(\theta _1\right)+\sin \left(\theta _1\right)
   \sqrt{z_{12}+1}\right) \sqrt{\gamma _1 \gamma
   _2+1}}{\sqrt{2}},
\end{equation}
\begin{displaymath}
z_{23}\to \frac{\left(\sin
   \left(\theta _1\right) \sqrt{z_{12}+1}-\cos
   \left(\theta _1\right) \sqrt{1-z_{12}}\right)
   \sqrt{\gamma _1 \gamma _2+1}}{\sqrt{2}},
\end{displaymath}
\begin{displaymath}
z_{14}\to
   \frac{\left(\sqrt{1-z_{12}} \cos \left(\theta
   _2\right)+\sin \left(\theta _2\right)
   \sqrt{z_{12}+1}\right) \sqrt{\gamma _1+\gamma
   _2}}{\sqrt{2} \sqrt{\gamma _2}},
\end{displaymath}
\begin{displaymath}
z_{24}\to
   \frac{\left(\sin \left(\theta _2\right)
   \sqrt{z_{12}+1}-\cos \left(\theta _2\right)
   \sqrt{1-z_{12}}\right) \sqrt{\gamma _1+\gamma
   _2}}{\sqrt{2} \sqrt{\gamma _2}},
\end{displaymath}
\begin{displaymath}
z_{34}\to
   Z_{34}-\frac{\cos \left(\theta _1-\theta _2\right)
   \sqrt{\gamma _1+\gamma _2} \sqrt{\gamma _1 \gamma
   _2+1}}{\sqrt{\gamma _2}} ,
\end{displaymath}
one is able to replace the conditions ((\ref{limits})-(\ref{limits2})) 
that the real two-qubit 
density matrix $\rho$ --- given by (\ref{BlooreDenMat}) --- must 
fulfill 
by
\begin{equation} \label{newlimits}
\gamma_{1} \in [0,1]; \hspace{.1in} \gamma_{2} \in [\gamma_{1},\frac{1}{\gamma_{1}}]; \hspace{.1in}
Z_{34} \in [-\gamma_{1}, \gamma_{1}]; \hspace{.1in} z_{12} \in [-1,1]; \hspace{.1in}
\theta_{1}, \theta_{2} \in [0, 2 \pi].
\end{equation}
We became aware of this set of transformations at a rather late stage of the research 
reported here, and have not been able so far --- somewhat disappointingly --- to exploit it in regards to
the HS separability-probability question. So, the results reported below
rely essentially upon the conditions ((\ref{limits})-(\ref{limits2})) and 
the parameterization in terms of the $z_{ij}$'s of Bloore..
\subsection{Previous analysis}
In \cite{slaterPRA2}, we studied the four nonnegativity 
conditions (as well as their counterparts  --- having 
completely parallel
cancellation and univariate function properties --- in the 15-dimensional
case of $4 \times 4$ density matrices with, in general, {\it complex} 
entries)  
using {\it 
numerical} (primarily quasi-Monte Carlo integration) methods. We found
a remarkably close fit to the function \cite[Figs. 3, 4]{slaterPRA2},
\begin{equation} \label{Freal}
\mathcal{S}_{real}(\nu) \approx  
\left(4+\frac{1}{5 \sqrt{2}}\right)
   B\left(\frac{1}{2},\sqrt{3}\right)^8
B_{\nu }\left(\frac{1}{2},\sqrt{3}\right),
\end{equation}
entering into our formula,
\begin{equation} \label{Vreal}
V^{HS}_{sep/real} = 2 \int_{0}^{1} \mathcal{J}_{real}(\nu) \mathcal{S}_{real}(\nu)
d \nu = \int_{0}^{\infty} \mathcal{J}_{real}(\nu) \mathcal{S}_{real}(\nu) d \nu,
\end{equation}
 for the 9-dimensional 
Hilbert-Schmidt {\it separable} volume
of the real $4 \times 4$ density matrices \cite[eq. (9)]{slaterPRA2}.
Here, $B$ denotes the (complete)
beta function, and $B_{\nu}$ the {\it incomplete} beta
function \cite{handbook},
\begin{equation}
B_{\nu}(a,b) =\int_{0}^{\nu} w^{a-1} (1-w)^{b-1} d w.
\end{equation}
Additionally \cite[eq. (10)]{slaterPRA2},
\begin{equation} \label{Jacreal}
\mathcal{J}_{real}(\nu) = \frac{\nu ^{3/2} \left(12 \left(\nu  (\nu +2) \left(\nu
   ^2+14 \nu +8\right)+1\right) \log \left(\sqrt{\nu
   }\right)-5 \left(5 \nu ^4+32 \nu ^3-32 \nu
   -5\right)\right)}{3780 (\nu -1)^9}
\end{equation}
is the 
(highly oscillatory near $\nu =1$ \cite[Fig.~1]{slaterPRA2}) 
jacobian function resulting from the transformation to the 
$\nu$ variable of the Bloore 
jacobian $(\Pi_{i=1}^{4} \rho_{ii})^{\frac{3}{2}}$.
(Perhaps we should refer to $\mathcal{J}_{real}(\nu)$ as a 
{\it marginal} jacobian, since it is the result of the integration 
of a {\it three}-dimensional jacobian function over two, say
$\rho_{11}$ and $\rho_{22}$, variables.)
\subsection{Computational limitations} \label{Strz}
Although we were able to implement the three (six-variable) 
nonnegativity conditions 
((\ref{firstcondition}), (\ref{secondcondition}) and (\ref{2by2}))
exactly in Mathematica in \cite{slaterPRA2}, 
for density matrices of the form (\ref{BlooreDenMat}),
we found that additionally incorporating the fourth
Peres-Horodecki (separability) one 
(\ref{Thirdcontion}) --- even
holding $\nu$ fixed at specific values --- seemed to yield a 
computationally intractable problem.
In fact, after the completion of \cite{slaterPRA2}, we consulted with
A. Strzebonski (the resident expert on these matters at 
Wolfram Inc.), and he wrote in regard to our problem that
``It looks like the [nine-dimensional four-condition 
separable real density matrix] 
problem is well out of range for CAD 
[the cylindrical algorithmic decomposition \cite{adam}]. The algorithm
is doubly exponential in the number of variables. Six variables is a lot
for CAD, so only very, very simple systems with six variables can be
solved. Adding one more inequality of total degree 4 makes a huge
difference. After an hour the algorithm is still in the projection
phase at 4 variables (it needs to go down to univariate polynomials)
and the projection polynomials already are huge: the last resultant
computed has degree 60 and 9520 terms, and the 3-variable projection
set already has 890 such polynomials...''
\subsection{Research design and objectives} 
In light of the apparent present computational intractability in
obtaining {\it exact} results in the
9-dimensional real 
(and {\it a fortiori} 15-dimensional complex) 
two-qubit cases, we adjusted the research 
program pursued in \cite{slaterPRA2}. We now sought to 
determine how
far we would have to curtail the dimension 
(the number of free parameters) of the two-qubit systems 
 in order to be 
able to obtain exact results using the same basic investigative framework.
Such results --- in addition to their own intrinsic interest --- might 
help us understand 
those previously obtained (basically 
numerically) 
in the {\it full} 9-dimensional real 
and 15-dimensional complex cases \cite{slaterPRA2}.

To pursue such a strategem, we chose to nullify various $m$-subsets of 
the six symmetrically-located 
off-diagonal pairs in the 9-parameter real 
density matrix (\ref{BlooreDenMat}), 
and tried to {\it exactly} implement the so-reduced 
non-negativity conditions
((\ref{firstcondition}), (\ref{secondcondition}), (\ref{2by2}) and 
(\ref{Thirdcontion})) --- both the first three (to obtain HS {\it total} 
volumes) and then all four 
jointly (to obtain HS {\it separable} volumes).
We leave the four {\it diagonal} entries themselves alone in all our 
analyses, so if we nullify
$m$ pairs of symmetically-located off-diagonal 
entries, we are left in a
(9-m)-dimensional setting.
We consider the various 
combinatorially distinct scenarios individually, though it would appear 
that we also could have grouped them into classes of scenarios equivalent under
{\it local} operations, and simply analyzed a single representative 
member of each equivalence class.

We will be examining a number of scenarios of various
 dimensionalities (that is, differing numbers of variables 
parameterizing $\rho$). 
In all of them,
we will seek to find the univariate
function $\mathcal{S}_{scenario}(\nu)$ 
(our primary computational and theoretical 
challenge) and the constant $c_{scenario}$, 
such that
\begin{equation} \label{Vsmall}
V^{HS}_{sep/scenario}= \int_{0}^{\infty} \mathcal{S}_{scenario}(\nu) \mathcal{J}_{scenario}(\nu) 
d \nu,
\end{equation}
and
\begin{equation} \label{Vbig}
V^{HS}_{tot/scenario}= c_{scenario} \int_{0}^{\infty}  \mathcal{J}_{scenario}(\nu) d \nu.
\end{equation}
Given such a pair of volumes, one can immediately
calculate the corresponding HS separability {\it probability},
\begin{equation}
P^{HS}_{sep/scenario}=\frac{V^{HS}_{sep/scenario}}{V^{HS}_{tot/scenario}}.
\end{equation}

Let us note that in the full 9-dimensional real and 15-dimensional
complex two-qubit cases 
recently studied in \cite{slaterPRA2}, it was quite natural
to expect 
that $\mathcal{S}_{real}(\nu)= \mathcal{S}_{real}(\frac{1}{\nu})$ (and
$\mathcal{S}_{complex}(\nu)= \mathcal{S}_{complex}(\frac{1}{\nu})$).
But, here, in our {\it lower}-dimensional scenarios, the nullification
of entries that we employ, breaks symmetry (duality), so we can not realistically
expect such a reciprocity property to hold, in general. 
Consequently, we adopt the more general, broader formula in  
(\ref{Vreal}) as our working formula (\ref{Vsmall}).
\section{Qubit-Qubit Analyses}
To begin, let us make the simple observation that since
the partial transposition operation on a $4 \times 4$ density matrix
interchanges only the (1,4) and (2,3) entries (and the (4,1) and (3,2) 
entries), any scenario which does not involve at least one of these
entries must only yield separable states.
\subsection{Five nullified pairs of off-diagonal entries --- 6 scenarios}
\subsubsection{4-dimensional real case --- $P^{HS}_{sep} = 
\frac{3 \pi}{16}$} \label{subsec4dim}
There are, of course, six ways of nullifying {\it five} 
of the six off-diagonal pairs of entries
of $\rho$. Of these, only two of the six yield  any non-separable
(entangled) states. In the four trivial 
(fully separable) scenarios, the lower-dimensional counterpart to
$\mathcal{S}_{real}(\nu)$ was of the form $\mathcal{S}_{scenario}(\nu)= 
c_{scenario}=2$.

In one of the two non-trivial scenarios, having the (2,3) and (3,2) 
pair of entries 
of $\rho$ left intact (not nullified), 
the separability function was
\begin{equation} \label{equationA}
\mathcal{S}_{[(2,3)]}(\nu) =
\begin{cases}
 2 \sqrt{\nu } & 0\leq \nu \leq 1 \\
 2 & \nu >1
\end{cases}.
\end{equation}
(It is of interest to note that 
$B_{\nu}(\frac{1}{2},1) = 
2 \sqrt{\nu}$, while in \cite{slaterPRA2}, we had conjectured
that $\mathcal{S}_{real}(\nu) \propto B_{\nu}(\frac{1}{2},\sqrt{3})$ 
and $\mathcal{S}_{complex}(\nu) \propto B_{\nu}(\frac{2 \sqrt{6}}{5},\frac{\sqrt{3}}{\sqrt{2}})$.)

In the other non-trivial scenario, 
with the (1,4) and (4,1) pair being the one not nullified, the 
separability function was --- in a dual manner (mapping $f(\nu)$ 
for $\nu \in [0,1]$ into $f(\frac{1}{\nu})$ for $\nu 
\in [1,\infty]$) --- equal to 
\begin{equation} \label{equationB}
\mathcal{S}_{[(1,4)]}(\nu) =
\begin{cases}
 2 & 0\leq \nu \leq 1 \\
 \frac{2}{\sqrt{\nu }} & \nu >1
\end{cases}.
\end{equation}
In both of these scenarios (having $c_{scenario} =2$)
for the total (separable and non-separable) HS 
volume, we obtained 
$V^{HS}_{tot} = \frac{\pi}{48} \approx 0.0654498$
and $V^{HS}_{sep} = \frac{\pi^2}{256} \approx .0385531$.
The corresponding
HS separability probability for the two non-trivial (dual) scenarios 
is, then, $\frac{3 \pi}{16} \approx 0.589049$.
\subsubsection{5-dimensional complex case --- $P^{HS}_{sep} 
= \frac{1}{3}$} 
\label{complex1}
Although our study here was initially intended to concentrate only on 
$4 \times 4$ density matrices with solely {\it real} entries, at a later
point in our analyses, we returned to (first) 
the $m=5$ case, but now with the single
non-nullified pair of symmetrically-located entries being {\it 
complex} in nature (so, obviously we have five variables/parameters 
{\it in toto} 
to consider, rather than four).

Again, we have only the same two scenarios (of the six 
combinatorially possible) being
separably non-trivial. Based on the (2,3) and (3,2) pair of entries,
the relevant function (with the slight change of notation to indicate 
complex entries) was
\begin{equation}
\mathcal{S}_{[\tilde{(2,3)}]}(\nu)= 
\begin{cases}
 \pi  \nu  & 0\leq \nu \leq 1 \\
 \pi  & \nu >1
\end{cases}
\end{equation}
and, dually, 
\begin{equation}
\mathcal{S}_{[\tilde{(1,4]})]}(\nu)= 
\begin{cases}
 \pi  & 0\leq \nu \leq 1 \\
 \frac{\pi }{\nu } & \nu >1
\end{cases}.
\end{equation}
So, the function $\sqrt{\nu}$, which appeared 
((\ref{equationA}), (\ref{equationB})) in the 
corresponding scenarios 
restricted to real entries, is replaced by $\nu$ itself in the
complex counterpart. (We note that $B_{\nu}(1,1) = \nu$.)

For both of these complex 
scenarios, we had $V^{HS}_{tot}= \frac{\pi}{120}$ and
$V^{HS}_{sep}= \frac{\pi}{360}$, for a 
particularly simple HS separability probability
of $\frac{1}{3}$.
\subsubsection{7-dimensional quaternionic  case --- $P^{HS}_{sep}
= \frac{1}{10}$}
Here we allow the single pair of non-null off-diagonal entries to be
{\it quaternionic} in nature \cite{asher2,adler} \cite[sec. IV]{batle}.
We found
\begin{equation}
\mathcal{S}_{[\widetilde{(2,3)}]}(\nu)=
\begin{cases}
 \frac{\pi^2  \nu^2}{2}  & 0\leq \nu \leq 1 \\
 \frac{\pi^2}{2}  & \nu >1
\end{cases}
\end{equation}
and, dually,
\begin{equation}
\mathcal{S}_{[\widetilde{(1,4]})]}(\nu)=
\begin{cases}
 \frac{\pi^2}{2}   & 0\leq \nu \leq 1 \\
 \frac{\pi^2 }{2 \nu^2 } & \nu >1
\end{cases}.
\end{equation}
(We note that $B_{\nu}(2,1) = \frac{\nu^2}{2}$.)
For both scenarios, we had $V^{HS}_{tot}=\frac{\pi^2}{2520}, V^{HS}_{sep}=\frac{\pi^2}{25200}$, giving us
$P^{HS}_{sep}= \frac{1}{10}$ --- which is the {\it smallest} 
probability we will report in this entire paper.

So, in our first set of simple ($m=5$) scenarios, we observe a decrease
in the probabilities of separability from the real to the complex to the 
quaternionic case, as well as a progression 
from $\sqrt{\nu}$ to $\nu$ to
$\nu^2$ in the functional forms occurring in the corresponding HS
separability probability functions.

The exponents of $\nu$ in this progression, that is $\frac{1}{2},1,2$
bear an evident relation to the {\it Dyson indices} \cite{dyson}, 1, 2, 4,
corresponding to the Gaussian orthogonal, unitary and symplectic ensembles 
\cite{desrosiers}. Certainly, this observation bears further investigation,
in particular since the foundational work of \.Zyczkowski and Sommers
\cite{szHS}) in computing the HS (separable {\it plus} nonseparable) volumes 
itself 
relies strongly on random matrix theory. 
(``[T]hese explicit results may be applied for estimation of the volume 
of the set of {\it entangled} [emphasis added] states'' 
\cite[p. 10125]{szHS}.) 
However, this theory is framed
in terms of the eigenvalues and eigenvectors of random matrices --- which
do not appear explicitly in the Bloore parameterization --- so, it is 
not altogether transparent in what manner one might proceed.
(But for the $m=5$ highly sparse density matrices for this set of scenarios,
one can explicitly transform between the eigenvalues and the Bloore parameters.)

\subsection{Four nullified pairs of off-diagonal entries --- 15 scenarios}
\subsubsection{5-dimensional real case --- $P^{HS}_{sep} = 
\frac{5}{8}; \frac{16}{3 \pi^2}$} \label{secd=5}
Here, there are fifteen possible scenarios, all with $V^{HS}_{tot} = 
\frac{\pi^2}{480}$. Six of them are trivial
(separability probabilities of 1), in which $c_{scenario}$
is either $\pi$ (scenarios [(1,2), (1,3)], [(1,2), (2,4)], [(1,3), (3,4)] and 
[(2,4), (3,4)]) or 4 (scenarios [(1,2), (3,4)] and [(1,3), (2,4)]).
Eight of the nine non-trivial scenarios 
all have --- similarly to the 4-dimensional analyses 
(sec.~\ref{subsec4dim}) ---- separability functions $\mathcal{S}(\nu)$ 
either of the form, 
\begin{equation}
\mathcal{S}_{scenario}(\nu) = \begin{cases}
 \pi  \sqrt{\nu } & 0\leq \nu \leq 1 \\
 \pi  & \nu >1
\end{cases},
\end{equation}
(for scenarios [(1,2), (2,3)], [(1,3), (2,3)], [(2,3), (2,4)] and
[(2,3), (3,4)])
 or, dually, 
\begin{equation}
\mathcal{S}_{scenario}(\nu) = \begin{cases}
 \pi  & 0\leq \nu \leq 1 \\
 \frac{\pi }{\sqrt{\nu }} & \nu >1
\end{cases}
\end{equation}
(for scenarios [(1,2), (1,4)], [(1,3), (1,4)], [(1,4), (2,4)] and
[(1,4), (3,4)]).
The corresponding HS
separability probabilities, for {\it 
all} eight of these non-trivial scenarios, 
are equal to 
$\frac{5}{8} = 0.625$. This result was, in all the eight cases, 
computed by taking the
the ratio of $V^{HS}_{sep} = \frac{\pi^2}{768}$ to $V^{HS}_{tot} = \frac{\pi^2}{480}$.

In the remaining (ninth) non-trivially entangled case --- based on the
non-nullified dyad 
[(1,4),(2,3)] --- we have, taking the ratio of $V^{HS}_{sep}= \frac{1}{90}$ to
$V^{HS}_{tot} = \frac{\pi^2}{480}$, a 
quite different Hilbert-Schmidt separability probability of 
$\frac{16}{3 \pi^2} \approx 0.54038$.
This isolated scenario (with $c_{scenario}=4$) 
can also be distinguished from the other eight 
partially entangled scenarios, in that it is the only one for
which entanglement occurs for {\it both} $\nu<1$ and $\nu>1$.
We have
\begin{equation} \label{suggestion}
\mathcal{S}_{[(1,4),(2,3)]}(\nu) = \begin{cases}
 4 \sqrt{\nu } & 0\leq \nu \leq 1 \\
 \frac{4}{\sqrt{\nu }} & \nu >1
\end{cases}.
\end{equation}
By way of illustration, in this specific case, we  have 
the scenario-specific marginal jacobian function,
\begin{equation}
\mathcal{J}_{[(1,4),(2,3)]}(\nu) = 
-\frac{\sqrt{\nu } \left(-3 \nu ^2+(\nu  (\nu +4)+1) \log
   (\nu )+3\right)}{30 (\nu -1)^5}.
\end{equation}
\subsubsection{6-dimensional {\it mixed} (real and complex) case ---$P^{HS}_{sep} = \frac{105 \pi}{512}; \frac{135 \pi}{1024}; \frac{3}{8}$}
Here, we again nullify all but two of the off-diagonal entries ($m=4$) of 
$\rho$, 
but allow the {\it first} of the two non-nullified entries to be
{\it complex} in nature. 
Making (apparently necessary) use of the
circular/trigonometric  transformation
$\rho_{11} = r^2 \sin{\theta}^2, \rho_{22} = r^2 \cos{\theta}^2$, we were
able to obtain an interesting variety of exact results.
One of these takes the form,
\begin{equation}
\mathcal{S}_{[\tilde{(1,2)},(1,4)]}(\nu)=
\mathcal{S}_{[\tilde{(1,3)},(1,4)]}(\nu)=
\begin{cases}
 \left\{\frac{4 \pi }{3},0\leq \nu \leq 1\right\} &
   \left\{\frac{4 \pi }{3 \sqrt{\nu }},\nu >1\right\}
\end{cases}.
\end{equation}
Now, we have $V^{HS}_{tot}= \frac{\pi^2}{1440}$ and
$V^{HS}_{sep} = \frac{7 \pi^3}{49152}$, so
$P^{HS}_{sep}= \frac{105 \pi}{512} \approx 0.644272$.
The two dual scenarios --- having the same three results --- are 
$[\tilde{(1,2)},(2,3)]$ and 
$[\tilde{(1,3)},(2,3)]$.

Additionally, we have an isolated scenario,
\begin{equation} \label{firstisolated}
\mathcal{S}_{[\tilde{(1,4)},(2,3)]}(\nu) =  \begin{cases}
 \left\{2 \pi  \sqrt{\nu },0\leq \nu \leq 1\right\} &
   \left\{\frac{2 \pi }{\nu },\nu >1\right\}
\end{cases},
\end{equation}
for which, $V^{HS}_{tot} = \frac{\pi^2}{1440}$ and
$V^{HS}_{sep}= \frac{3 \pi^3}{32768}$, so
$P^{HS}_{sep}= \frac{135 \pi}{1024} \approx 0.414175$.
(Note the presence of {\it both} $\sqrt{\nu}$ and $\nu$ in (\ref{firstisolated}) --- apparently related to the mixed [real and complex] nature of this 
scenario (cf. (\ref{secondmixed})).)

Further.
\begin{equation}
\mathcal{S}_{[\tilde{(1,4)},(2,4)]}(\nu)=
\mathcal{S}_{[\tilde{(1,4)},(3,4)]}(\nu)= \begin{cases}
 \left\{\frac{4 \pi }{3},0\leq \nu \leq 1\right\} &
   \left\{\frac{4 \pi }{3 \nu },\nu >1\right\}
\end{cases},
\end{equation}
the dual scenarios being $[\tilde{(2,3)},(2,4)]$ and $[\tilde{(2,3)},(3,4)]$.
For all four of these scenarios, $V^{HS}_{tot}= \frac{\pi^2}{1440}$ and
$V^{HS}_{sep}= \frac{\pi^2}{3840}$,
so $P^{HS}_{sep} =\frac{3}{8} =0.375$.
\subsubsection{7-dimensional complex
case --- $P^{HS}_{sep} = \frac{2}{5}$}
Here, in an $m=4$ setting, 
we nullify four of the six off-diagonal pairs of the $4 \times 4$ 
density matrix, allowing the remaining two pairs {\it both} to be complex.
We have (again observing a shift from $\sqrt{\nu}$ in the real case
to $\nu$ in the complex case)
\begin{equation} \label{complex7}
\mathcal{S}_{[\tilde{(1,2)},\tilde{(1,4)}]}(\nu) =\mathcal{S}_{[\tilde{(1,3)},\tilde{(1,4)}]}(\nu)= 
\mathcal{S}_{[\tilde{(1,4)},\tilde{(2,4)}]}(\nu)=
\mathcal{S}_{[\tilde{(1,4)},\tilde{(3,4)}]}(\nu) =
\begin{cases}
 \frac{\pi ^2}{2} & 0\leq \nu \leq 1 \\
 \frac{\pi ^2}{2 \nu } & \nu >1
\end{cases}. 
\end{equation}
Since $V^{HS}_{tot}= \frac{\pi^2}{5040}$ and
$V^{HS}_{sep}= \frac{\pi^2}{12600}$, we have
$P^{HS}_{sep}= \frac{2}{5} = 0.4$.
We have the same three outcomes for the four dual 
scenarios $[\tilde{(1,2)},\tilde{(2,3)}],
[\tilde{(1,3)},\tilde{(2,3)}],
[\tilde{(2,3)},\tilde{(2,4)}]$ and 
$[\tilde{(2,3)},\tilde{(3,4)}]$, as well as --- rather remarkably --- for 
the (again isolated 
[cf. (\ref{firstisolated})])
scenario $[\tilde{(1,4)},\tilde{(2,3)}]$, having the (somewhat different)
separability function (manifesting entanglement for both
$\nu<1$ and $\nu > 1$),
\begin{equation} \label{secondmixed}
\mathcal{S}_{[\tilde{(1,4)},\tilde{(2,3)}]}(\nu)  = 
\begin{cases}
 \pi ^2 \nu & 0\leq \nu \leq 1 \\
 \frac{\pi ^2}{\nu } & \nu >1
\end{cases}.
\end{equation}
(However, $c_{scenario} = \pi^2$ for this isolated scenario, while it
equals
$\frac{\pi^2}{2}$ for the other eight.)
The remaining six (fully separable) scenarios (of the fifteen possible)
simply have $P^{HS}_{sep}=1$.
\subsubsection{8-dimensional mixed (real and quaternionic) case}
We report here that
\begin{equation}
c_{[\widetilde{(1,2)},(1,4)]} =\frac{8 \pi^2}{15}, \hspace{.2in} 
c_{[(1,2),\widetilde{(1,4)}]}= 32,
\end{equation}
where as before the wide tilde notation denote the quaternionic 
off-diagonal entry.
\subsection{Three nullified pairs of off-diagonal entries --- 20 scenarios}
\subsubsection{6-dimensional 
real case --- $P^{HS}_{sep} =2 -\frac{435 \pi}{1024}; \approx \frac{9}{16}$}
Here ($m=3$), 
there are twenty possible scenarios --- nullifying {\it triads} of
off-diagonal pairs in $\rho$.
Of these twenty, there are 
four totally separable scenarios --- corresponding to the non-nullified triads [(1,2), (1,3), (2,4)], [(1,2), (1,3), (3,4)], [(1,2), (2,4), (3,4)] and [(1,3), (2,4), (3,4)] --- with $c_{scenario} = \frac{\pi^2}{2}$ and 
$V^{HS}_{tot} = V^{HS}_{sep} = \frac{\pi^3}{5760}$.
To proceed further in this 6-dimensional case --- in which we 
began to 
encounter some computational difficulties --- we sought, again,  to 
enforce the four  nonnegativity conditions
((\ref{firstcondition}), (\ref{secondcondition}), (\ref{2by2}),
(\ref{Thirdcontion})), but only after setting $\nu$ to specific values,
rather than allowing $\nu$ to vary.
We chose the nine values
$\nu= \frac{1}{5}, \frac{2}{5}, \frac{3}{5}, \frac{4}{5}$, 1, 2, 3, 4 and 5.
Two of the scenarios (with the triads [(1,2), (2,3), (3,4)] and
[(1,3),(2,3),(2,4)]) could, then, be seen to fit unequivocally  
into our earlier observed 
predominant pattern,
having the piecewise separability function,
\begin{equation} \label{1case}
\mathcal{S}_{[(1,2), (2,3), (3,4)]}(\nu) = 
\mathcal{S}_{[(1,3), (2,3), (2,4)]}(\nu) =
\begin{cases}
 \frac{\pi ^2 \sqrt{\nu }}{2} & 0\leq \nu \leq 1 \\
 \frac{\pi ^2}{2 } & \nu >1
\end{cases}.
\end{equation}
We, then, computed for these two scenarios that 
$V^{HS}_{tot} = \frac{\pi^3}{5760} \approx 0.00538303$ and 
(again making use of the transformation
$\rho_{11} = r^2 \sin{\theta}^2, \rho_{22} = r^2 \cos{\theta}^2$) that 
$V^{HS}_{sep} = 2 \left(\frac{\pi ^3}{5760}-\frac{29 \pi^4}{786432}\right) 
\approx 0.00358207$. This gives  us 
$P^{HS}_{sep} = 2-\frac{435 \pi }{1024} \approx 0.665437$.
For two dual dyads, we have the same volumes and separability 
probability and, now, the piecewise separability function,
\begin{equation} \label{onecase}
\mathcal{S}_{[(1,2), (1,4), (3,4)]}(\nu)= \mathcal{S}_{[(1,3), (1,4), (2,4)]}(\nu) =
\begin{cases}
 \frac{\pi ^2}{2 } & 0\leq \nu \leq 1 \\
 \frac{\pi ^2}{2 \sqrt{\nu}} & \nu >1
\end{cases}.
\end{equation}

\begin{figure}
\includegraphics{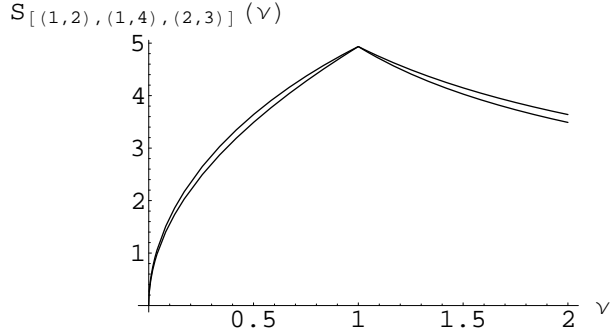}
\caption{\label{fig:good}Plot of $\mathcal{S}_{[(1,2),(1,4),(2,3)]}(\nu)$ and the
close fitting function  (\ref{newcase}).
The former function slightly dominates the 
latter except at $\nu=1$, where they both equal $\frac{\pi^2}{2} 
\approx 4.9348$.}
\end{figure}

Additionally, in Fig.~\ref{fig:good}, 
we are able to plot 
$\mathcal{S}_{[(1,2),(1,4),(2,3)]}(\nu)$ along with its close fit 
to
\begin{equation} \label{newcase}
\mathcal{S}_{fit}(\nu) =
\begin{cases}
 \frac{\pi ^2 \sqrt{\nu }}{2} & 0\leq \nu \leq 1 \\
 \frac{\pi ^2}{2 \sqrt{\nu }} & \nu >1
\end{cases}.
\end{equation}
(The analogous plots for the scenarios [(1,3), (1,4), (2,3)] and 
[(1,2), (1,3), (2,3)] appear to be precisely the same in character as 
Fig.~\ref{fig:good}.)
If we use the close fit (\ref{newcase}) as a proxy for
$\mathcal{S}_{[(1,2),(1,4),(2,3)]}(\nu)$, we obtain an 
{\it approximate} HS separability
probability of $\frac{9}{16} =\frac{\frac{\pi^3}{10240}}{\frac{\pi^3}{5760}}
=0.5625$.

We have not, to this point, been able to explicitly 
and succinctly characterize
the functions $\mathcal{S}_{scenario}(\nu)$ for non-trivial 
fully real $m=3$ scenarios 
other than the dual pair 
((\ref{1case}), (\ref{onecase})).

In all the separably non-trivial scenarios so far presented and discussed,
we have had the relationship $\mathcal{S}_{scenario}(1) = c_{scenario}$. However,
in our present $m=3$ setting (three pairs of 
nullified off-diagonal entries), we have
situations in which $\mathcal{S}_{scenario}(1) <  c_{scenario}$.
The values of $c_{scenario}$ in the sixteen non-trivial 
fully real $m=3$ scenarios 
are either $\frac{\pi^2}{2} \approx 4.9348$ (twelve occurrences) or
$\frac{4 \pi}{3} \approx 4.18879$ (four occurrences --- [(1,2), (1,3), (1,4)],  [(1,2), (2,3), (2,4)], [(1,3), (2,3), (3,4)] and [(1,4), (2,4), (3,4)]). 
In all four of 
the latter ($\frac{4 \pi}{3}$) occurrences, though, 
we have the {\it inequality},
\begin{equation}
\mathcal{S}_{scenario}(1) = 
\frac{1}{24} \left(12+16 \pi +3 \pi ^2\right)
\approx 3.8281 < \frac{4 \pi}{3} \approx 4.18879,
\end{equation}
as well as a parallel 
inequality for four of the twelve former ($\frac{\pi^2}{2}$) 
cases.
The implication of these inequalities for those eight 
scenarios is that at $\nu=1$ (the value associated with the fully
mixed [separable] classical state), 
that is, when $\rho_{11} \rho_{44} = \rho_{22} \rho_{33}$, there do exist
non-separable states.

\subsubsection{7-dimensional mixed (one complex and two real) case --- 
$P_{sep}^{HS} = \frac{11}{16}$}
Here, in an $m=3$ setting, we take the first entry of the 
non-nullified triad to be 
complex and the other two real. 
Of the twenty possible scenarios, four ---- $[\tilde{(1,2)},(1,3),(1,4)], 
[\tilde{(1,2)},(2,3),(2,4)],
[\tilde{(1,3)},(2,3),(3,4)]$ and 
$[\tilde{(1,4)},(2,4),(3,4)]$ --- had $c_{scenario}=\frac{\pi^2}{2} 
\approx 4.9348$ and these four all had the same (lesser) value of
\begin{equation} \label{Sscenario}
\mathcal{S}_{scenario}(1) = \frac{56}{27}+\frac{\pi^2}{4} \approx 4.54148.
\end{equation}

There were seven scenarios with $c_{scenario} = \frac{16 \pi}{9} 
\approx 5.58505$. Three of them --- $[\tilde{(1,2)},(1,3),(2,4)],
[\tilde{(1,2)},(1,4),(2,3)]$ and 
$[\tilde{(1,3)},(1,4),(2,3)]$  --- had
$\mathcal{S}_{scenario}(1)=  
\frac{16 \pi}{9}$ (manifesting equality), 
while four --- $[\tilde{(1,2)},(1,3),(2,3)],
[\tilde{(1,2)},(1,4),(2,4)],
[\tilde{(1,3)},(1,4),(3,4)]$ and
$[\tilde{(2,3)},(2,4),(3,4)]$ --- had the result 
(\ref{Sscenario}) (manifesting inequality).

The remaining nine of the twenty scenarios all had $c_{scenario} = 
\mathcal{S}_{scenario}(1) = \frac{2 \pi^2}{3} \approx 6.57974$.
For one of them, we obtained 
\begin{equation}
\mathcal{S}_{[\tilde{(1,2)},(2,3),(3,4)]}(\nu) = \begin{cases}
 \frac{2 \pi ^2}{3} & \nu \geq 1 \\
 \frac{2 \pi ^2 \sqrt{\nu }}{3} & 0<\nu <1
\end{cases},
\end{equation}
with associated values of $V_{tot}^{HS}= \frac{\pi^3}{20160}$,
$V_{sep}^{HS}= \frac{11 \pi^3}{322560}$ and
$P_{sep}^{HS}= \frac{11}{16} \approx 0.6875$.
A dual scenario to this one that we were able to find was 
$[\tilde{(1,2)},(1,4),(3,4)]$.
The separability functions --- and, hence, separability 
probabilities --- for the other eighteen scenarios, however, are unknown
to us at present.
\subsubsection{8-dimensional mixed (two complex and one real)  case}
Our sole result in this category is
\begin{equation}
c_{[\tilde{(1,2)},\tilde{(1,3)},(1,4)]}=\frac{8 \pi^2}{15}.
\end{equation}
\subsubsection{9-dimensional complex case}
Now, we have three off-diagonal complex entries, requiring {\it six}
parameters for their specification. According to the remarks of
Strzebonski (sec.~\ref{Strz}), this is about the limit in the number 
of free off-diagonal parameters for which we might
hopefully be able to determine associated separability functions.

As initial findings, we obtained
\begin{equation}
\mathcal{S}_{[\tilde{(1,4)},\tilde{(2,3)},\tilde{(2,4)}]}(1) = 
c_{[\tilde{(1,4)},\tilde{(2,3)},\tilde{(2,4)}]} =
\frac{\pi^3}{4},
\end{equation}
and also for scenarios $[\tilde{(1,2)},\tilde{(1,4)},\tilde{(3,4)}], 
[\tilde{(1,3)},\tilde{(1,4)},\tilde{(2,4)}]$ and
$[\tilde{(1,4)},\tilde{(2,3)},\tilde{(3,4)}]$,
while
\begin{equation}
c_{[\tilde{(1,2)},\tilde{(1,3)},\tilde{(1,4)}]} = 
c_{[\tilde{(1,3)},\tilde{(2,3)},\tilde{(3,4)}]} =
c_{[\tilde{(1,4)},\tilde{(2,3)},\tilde{(2,4)}]} =
\frac{\pi^3}{6}.
\end{equation}
\subsection{Two or fewer nullified pairs of off-diagonal entries}
\subsubsection{7-dimensional real case}
The [(1,2), (1,3), (2,4), (3,4)] scenario is
the only fully separable one
of the fifteen possible ($m=2$). For all 
the other fourteen non-trivial scenarios,
there are non-separable states {\it both} for $\nu<1$ and $\nu>1$.
For all fifteen scenarios, we have $c_{scenario} = \frac{2 \pi^2}{3} 
\approx 6.57974$.
Otherwise, we 
have not so far been able to extend the analyses above to this $m=2$ 
fully real
case (and {\it a fortiori} the $m=1$ fully real case), 
even to determine specific values of $\mathcal{S}_{scenario}(1)$.
\subsubsection{8-dimensional real case} \label{sec8dim}
Here we have $c_{scenario} = \frac{8 \pi^2}{9} \approx 
8.77298$ for all the six possible (separably non-trivial)
scenarios ($m=1$). Let us note that this is, in terms of preceding values of 
these constants (for the successively lower-dimensional fully real 
scenarios), $\frac{8 \pi^2}{9} = \frac{4}{3} (\frac{2 \pi^2}{3})$, while
$\frac{2 \pi^2}{3} = \frac{4}{3} (\frac{\pi^2}{2})$. 
Also, 
$\frac{32 \pi^2}{27} = \frac{4}{3} (\frac{8 \pi^2}{9}$), the 
further relevance of
which will be apparent in relation to our discussion of 
the full 9-dimensional real scenario (sec.~\ref{concludingremarks}).

\section{Qubit-Qutrit Analyses} \label{qubitqutrit}
The cancellation property, we exploited above,  of the Bloore
parameterization --- by which the determinant and 
principal minors of density matrices
can be {\it factored} 
into products of (nonnegative) diagonal entries and terms just 
involving off-diagonal parameters ($z_{ij}$) --- clearly 
extends to $n \times n$ density matrices. It initially appeared to us that the advantage of the 
parameterization in studying the 
two-qubit HS separability probability question would diminish if one
were to examine the two-qubit separability problem for other (possibly 
{\it monotone}) metrics than the HS one, 
or even the qubit-{\it qutrit} {\it HS} separability
probability question. 
But upon some further analysis, we have found that the 
nonnegativity condition for the
determinant of the partial transpose of a real $6 \times 6$ (qubit-qutrit)
density matrix (cf. (\ref{firstcondition}))
can be expressed in terms of the corresponding $z_{ij}$'s
and {\it two} ratio variables (thus, not requiring the five independent
diagonal variables individually), 
\begin{equation} \label{tworatios}
\nu_{1}= \frac{\rho_{11} \rho_{55}}{\rho_{22} \rho_{44}}, \hspace{.2in}
\nu_{2}= \frac{\rho_{22} \rho_{66}}{\rho_{33} \rho_{55}},
\end{equation}
rather than simply one ($\nu$) as in the
$4 \times 4$ case. 
(We compute the qubit-qutrit 
partial transpose by transposing in place the four
$3 \times 3$ blocks of $\rho$, rather than --- as we might 
alternatively have done --- the nine $2 \times 2$ blocks.)
\subsection{Fourteen nullified pairs of off-diagonal entries --- 15 scenarios}
\subsubsection{6-dimensional real case --- $P^{HS}_{sep} =\frac{3 \pi}{16}$}
To begin our examination of  the qubit-qutrit case, we study the 
($m=14$) scenarios, 
in which only a single pair of real entries is left 
intact and all other off-diagonal pairs of the $6 \times 6$ 
density matrix are nullified. (We not only require that the determinant
of the partial transpose of $\rho$ be nonnegative 
for separability to hold --- as suffices 
in the qubit-qubit case, given that $\rho$ itself is a density 
matrix  \cite{ver,augusiak} --- but also, {\it per} the 
Sylvester criterion,  a nested series of 
principal leading minors of $\rho$.)
We have six separably non-trivial scenarios. 
(For all of them, $V_{tot}^{HS}= \frac{\pi}{1440}$.)

Firstly, we
have the separability
function,
\begin{equation} \label{S15}
\mathcal{S}_{[(1,5)]}^{6 \times 6}(\nu_{1})=
\begin{cases}
 2 & \nu _1\leq 1 \\
 \frac{2}{\sqrt{\nu _1}} & \nu_{1}>1
\end{cases}.
\end{equation}
The dual scenario to this is [(2,4)].
Further,
\begin{equation}
\mathcal{S}_{[(1,6)]}^{6 \times 6}(\nu_{1},\nu_{2})=
\begin{cases}
 2 & \nu_{1} \nu _2 \leq 1 \\ 
 \frac{2}{\sqrt{\nu _1 \nu _2}} & \nu _1 \nu_2 >  1 
\end{cases},
\end{equation}
with the dual scenario here being [(3,4)].
Finally,
\begin{equation}
\mathcal{S}_{[(2,6)]}^{6 \times 6}(\nu_{2})=
\begin{cases}
 2 & \nu _2\leq 1 \\
 \frac{2}{\sqrt{\nu _2}} & \nu_{2} > 1
\end{cases},
\end{equation}
having the dual [(3,5)].

The remaining nine possible scenarios --- the same as their 
complex counterparts in the immediate
next analysis --- are all fully separable in character.

We have found that $V^{HS}_{sep} = \frac{\pi^2}{7680}$ for the
six non-trivially separable scenarios here, so $P^{HS}_{sep} = 
\frac{3 \pi}{16} \approx 0.589049$, as in the qubit-qubit analogous
case (sec.~\ref{subsec4dim}).
\subsubsection{7-dimensional complex case --- $P^{HS}_{sep}= \frac{1}{3}$}  \label{ComplexSection}
Now, we allow the single non-nullified pair of off-diagonal entries
to be complex in nature (the two paired entries, of course, being complex
conjugates of one another). ($V_{tot}^{HS}= \frac{\pi}{5040}$ for this
series of fifteenb scenarios.)
Then, we have (its dual being $[\tilde{(2,4)}]$)
\begin{equation}
\mathcal{S}_{[\tilde{(1,5)}]}^{6 \times 6}(\nu_{1})=
\begin{cases}
 \pi  & \nu _1\leq 1 \\
 \frac{\pi }{\nu _1} & \nu_{1} > 1
\end{cases}.
\end{equation}

Further, we have  (with the dual $[\tilde{(3,4)}]$)
\begin{equation}
\mathcal{S}_{[\tilde{(1,6)}]}^{6 \times 6}(\nu_{1},\nu_{2})=
\begin{cases}
 \pi  & \nu_{1} \nu _2 \leq 1 \\ 
 \frac{\pi }{\nu _1 \nu _2} & \nu _1 \nu_{2} > 1
\end{cases}
\end{equation}
and (its dual being $[\tilde{(3,5)}]$),
\begin{equation}
\mathcal{S}_{[\tilde{(2,6)}]}^{6 \times 6}(\nu_{2})=
\begin{cases}
 \pi  & \nu _2\leq 1 \\
 \frac{\pi }{\nu _2} & \nu_{2} > 1.
\end{cases}.
\end{equation}
For all six of these scenarios,
$V_{sep}= \frac{\pi}{15120}$, so $P_{HS}^{sep} = \frac{1}{3}$.
\subsection{Thirteen nullified pairs of off-diagonal entries --- 105 scenarios}
\subsubsection{7-dimensional real case --- $P^{HS}_{sep}=\frac{5}{8}; \frac{5}{16}; \frac{3 \pi}{32}; \frac{16}{3 \pi^2}$}
Continuing along similar lines ($m=13$), we have 105 combinatorially
distinct possible scenarios. 
Among the separably non-trivial scenarios, we have
\begin{equation}
\mathcal{S}_{[(1,2),(1,5)]}^{6 \times 6}(\nu_1)=
\mathcal{S}_{[(1,4),(1,5)]}^{6 \times 6}(\nu_1)=
\begin{cases}
 \pi  & \nu _1\leq 1 \\
 \frac{\pi }{\sqrt{\nu _1}} & \nu_{1} > 1
\end{cases},
\end{equation}
(duals being [(1,2),(2,4)] and [(1,4),(2,4)]).  We computed  $V^{HS}_{tot}= 
\frac{\pi^2}{20610}, V^{HS}_{sep}= \frac{\pi^2}{32256}$, so
$P^{HS}_{sep}=\frac{5}{8} = 0.625$ for these scenarios.

Also,
\begin{equation}
\mathcal{S}_{[(1,3),(1,6)]}^{6 \times 6}(\nu_{1},\nu_{2}) =
\mathcal{S}_{[(1,4),(1,6)]}^{6 \times 6}(\nu_{1},\nu_{2}) =
\begin{cases}
 \pi  & \nu_{1} \nu_{2} < 1  \\
 \frac{\pi }{\sqrt{\nu _1} \sqrt{\nu _2}} & \nu_{1} \nu_{2} \geq  1
\end{cases}.
\end{equation}
We, then,  have $V^{HS}_{tot}=
\frac{\pi^2}{20610}, V^{HS}_{sep}= \frac{\pi^2}{64512}$, so
$P^{HS}_{sep}=\frac{5}{16} = 0.3125$.

Additionally,
\begin{equation}
\mathcal{S}_{[(1,4),(2,6)]}^{6 \times 6}(\nu_{2}) =
\begin{cases}
 4 & \nu _2\leq 1 \\
 \frac{4}{\sqrt{\nu _2}} & \nu_{2} > 1.
\end{cases}.
\end{equation}
For this scenario, we have 
$V^{HS}_{tot}=
\frac{\pi^2}{20610}, V^{HS}_{sep}= \frac{\pi^2}{215040}$, so
$P^{HS}_{sep}=\frac{3 \pi}{32} \approx 0.294524$.

Further still,
\begin{equation}
\mathcal{S}_{[(1,5),(2,4)]}^{6 \times 6}(\nu_{1}) =
\begin{cases}
 \frac{4}{\sqrt{\nu _1}} & \nu _1>1 \\
 4 \sqrt{\nu _1} & \nu_{1} \leq 1
\end{cases}.
\end{equation}
For this scenario, we have
$V^{HS}_{tot}=
\frac{\pi^2}{20610}, V^{HS}_{sep}= \frac{1}{3780}$, so
$P^{HS}_{sep}=\frac{16}{3 \pi^2} \approx 0.54038$. 

Further,
\begin{equation}
\mathcal{S}_{[(1,2),(2,6)]}^{6 \times 6}(\nu_{2})=
\begin{cases}
 \pi  & \nu _2\leq 1 \\
 2 \left(\cos ^{-1}\left(\sqrt{1-\frac{1}{\nu
   _2}}\right)+\frac{\sqrt{\nu _2-1}}{\nu _2}\right) &
   \nu_{2} > 1
\end{cases}.
\end{equation}
The separability function for [(1,3),(1,5)] is obtained from this one
by replacing $\nu_{2}$ by $\nu_{1}$.

Also,
\begin{equation}
\mathcal{S}^{6 \times 6}_{[(1,2),(3,4)]}(\nu_{1},\nu_2) =
\end{equation}
\begin{displaymath}
\begin{cases}
\pi \sqrt{\nu_{1}} \sqrt{\nu_{2}} & \nu_{1} \nu_{2} <1 \\
 4 \sqrt{1-\frac{1}{\nu _1 \nu _2}}-\frac{2 \left(i \log
   \left(\frac{\sqrt{\nu _1 \nu _2-1}+i}{\sqrt{\nu _1}
   \sqrt{\nu _2}}\right) \nu _1 \nu _2+\sqrt{\nu _1 \nu
   _2-1}\right)}{\sqrt{\nu _1} \sqrt{\nu _2}} & \nu_{1} \nu_{2} \geq 1
\end{cases}.
\end{displaymath}
The separability function for [(1,2),(3,5)] 
can be obtained from this one by setting $\nu_{1}=1$.
\subsubsection{8-dimensional mixed (real and complex) case --- $P^{HS}_{sep} = \frac{105 \pi}{512}$}
Further, we have (with $V_{tot}^{HS} = 
\frac{\pi^2}{80640}$ for all scenarios),
\begin{equation} \label{firstmixed}
\mathcal{S}_{[\tilde{(1,2)},(2,4)]}^{6 \times 6}(\nu_{1})=
\mathcal{S}_{[\tilde{(1,4)},(2,4)]}^{6 \times 6}(\nu_{1})=
\begin{cases}
 \frac{4 \pi }{3} & \nu _1\geq 1  \\
 \frac{4 \pi  \sqrt{\nu _1}}{3} & 0<\nu _1<1
\end{cases}.
\end{equation}
Since $V^{HS}_{sep} = \frac{\pi^3}{393216}$, we have
$P^{HS}_{sep} = \frac{105 \pi}{512} \approx 0.644272$ for both 
these scenarios.

Further,
\begin{equation}
\mathcal{S}_{[\tilde{(1,3)},(2,4)]}^{6 \times 6}(\nu_{1})=
\begin{cases}
 \frac{4 \pi  \sqrt{\nu _1}}{3} & 0<\nu _1 \leq 1 \\
 2 \pi -\frac{2 \pi }{3 \nu _1} & \nu _1>1
\end{cases}
\end{equation}
and
\begin{equation}
\mathcal{S}_{[\tilde{(1,3)},(3,4)]}^{6 \times 6}(\nu_{1},\nu_{2})=
\mathcal{S}_{[\tilde{(1,4)},(3,4)]}^{6 \times 6}(\nu_{1},\nu_{2})=
\begin{cases}
 \frac{4 \pi }{3} & \nu _1 \nu _2\geq 1 \\
 \frac{4}{3} \pi  \sqrt{\nu _1 \nu _2} & 0<\nu _1 \nu _2<1
\end{cases}.
\end{equation}

Additionally,
\begin{equation} \label{puzzle}
\mathcal{S}_{[\tilde{(2,3)},(3,4)]}^{6 \times 6}(\nu_{1},\nu_{2})= 
\begin{cases}
 \frac{4 \pi }{3} & \nu _1 \nu _2\geq 1 \\
 \frac{2}{3} \pi  \sqrt{\nu _1 \nu _2} \left(3-\nu _1 \nu_2  \right) & 0<\nu _1 \nu _2<1
\end{cases}
\end{equation}
and
\begin{equation}
\mathcal{S}_{[\tilde{(1,2)},(3,4)]}^{6 \times 6}(\nu_{1},\nu_{2})=
\begin{cases}
 \frac{4 \pi }{3} & \nu _1 \nu _2=1 \\
 \frac{4}{3} \pi  \sqrt{\nu _1 \nu _2} & 0<\nu _1 \nu _2<1
   \\
 \pi  \left(\sqrt{\nu _1 \nu _2}+2\right)-\frac{2 \pi }{3
   \nu _1 \nu _2} & \nu _1 \nu _2>1
\end{cases}.
\end{equation}

\begin{figure}
\includegraphics{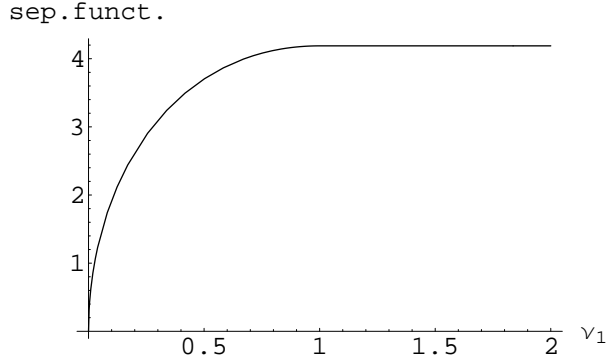}
\caption{\label{fig:S2324}Plot of the separability function
$S_{[\tilde{(2,3)},(2,4)]}^{6 \times 6}(\nu_{1})$}
\end{figure}

We have also obtained the separability function
(Fig.~\ref{fig:S2324})
\begin{equation}
\mathcal{S}_{[\tilde{(2,3)},(2,4)]}^{6 \times 6}(\nu_{1})=
\begin{cases}
 \frac{4 \pi }{3} & \nu _1\geq 1 \\
 \frac{2}{3} \pi  \left(3 -\nu _1 \right) \sqrt{\nu _1} &
   0<\nu _1<1
\end{cases}.
\end{equation}
Of the 105 possible scenarios, sixty had $S^{6 \times 6}_{scenario}(1,1) = 
c_{scenario} = \frac{4 \pi}{3}$, thirty-three 
had $S^{6 \times 6}_{scenario}(1,1) =
c_{scenario} = 2 \pi$, and twelve (for example, $[\tilde{(3,4)},(5,6)]$) had
$S^{6 \times 6}_{scenario}(1,1) = \frac{4 \pi}{3} < c_{scenario} = 2 \pi$.
\subsubsection{9-dimensional complex case --- $P^{HS}_{sep} = \frac{1}{3}; \frac{2}{5}$}
We have obtained the results
\begin{equation}
\mathcal{S}_{[\tilde{(1,2)},\tilde{(2,4)}]}^{6 \times 6}(\nu_{1})= 
\mathcal{S}_{[\tilde{(1,4)},\tilde{(2,4)}]}^{6 \times 6}(\nu_{1})=
\begin{cases}
 \frac{\pi ^2 \nu_{1} }{2} & 0\leq \nu_{1} \leq 1 \\
 \frac{\pi ^2}{2 } & \nu_{1} >1
\end{cases}.
\end{equation}
Since $V^{HS}_{tot} = \frac{\pi^2}{362880}$ and
$V^{HS}_{sep}= \frac{\pi^2}{907200}$, we have here
$P^{HS}_{sep} = \frac{2}{5} = 0.4$. 
We have the same three outcomes also based on the separability function,
\begin{equation}
\mathcal{S}_{[\tilde{(1,4)},\tilde{(3,4)}]}^{6 \times 6}(\nu_{1},\nu_{2})=
\begin{cases}
 \frac{\pi ^2}{2} & \nu _1 \nu _2\geq 1 \\
 \frac{1}{2} \pi ^2 \nu _1 \nu _2 & \nu _1 \nu _2<1
\end{cases}.
\end{equation}

Further,
\begin{equation}
\mathcal{S}_{[\tilde{(1,2)},\tilde{(3,4)}]}^{6 \times 6}(\nu_{1},\nu_{2})=
\begin{cases}
 \frac{2 \pi ^2 \nu _1 \nu _2-\pi ^2}{2 \nu _1 \nu _2} &
   \nu _1 \nu _2>1 \\
 \frac{1}{2} \pi ^2 \nu _1 \nu _2 & 0<\nu _1 \nu _2 \leq 1
\end{cases}.
\end{equation}
and
\begin{equation}
\mathcal{S}_{[\tilde{(1,3)},\tilde{(2,4)}]}^{6 \times 6}(\nu_{1})=
\begin{cases}
 \frac{\pi ^2 \nu _1}{2} & 0<\nu _1 \leq 1 \\
 \pi ^2-\frac{\pi ^2}{2 \nu _1} & \nu _1>1
\end{cases}.
\end{equation}
For both of these last two scenarios, we have $V_{tot}^{HS}= \frac{\pi^2}{362880}$ and 
$V^{HS}_{sep}= \frac{\pi^2}{362880}$, leading to
$P^{HS}_{sep}= \frac{1}{3} \approx 0.33333$.
Also, we have these same three outcomes based on the 
separability function,
\begin{equation}
\mathcal{S}_{[\tilde{(2,3)},\tilde{(3,4)}]}^{6 \times 6}(\nu_{1},\nu_{2})=
\begin{cases}
 \frac{\pi ^2}{2} & \nu _1 \nu _2\geq 1 \\
 \frac{1}{2} \pi ^2 \nu _1 \nu _2 \left(2 - \nu _1 \nu
   _2\right) & 0<\nu _1 \nu _2<1
\end{cases}.
\end{equation}

Of the 105 possible scenarios --- in complete parallel to those 
in the immediately preceding section --- sixty 
had $S^{6 \times 6}_{scenario}(1,1) =
c_{scenario} = \frac{\pi^2}{2}$, thirty-three
had $S^{6 \times 6}_{scenario}(1,1) =
c_{scenario} = \pi^2$, and twelve (for example, 
$[\tilde{(3,4)},\tilde{(5,6)}]$) had
$S^{6 \times 6}_{scenario}(1,1) = \frac{\pi^2}{2} < c_{scenario} =  \pi^2$.

Our results in this (9-dimensional) section and the 
(8-dimensional) one immediately preceding it are still incomplete 
with respect to various scenario-specific separability functions
and, thus, the associated HS separability properties.
\subsection{Twelve nullified pairs of off-diagonal entries --- 455 scenarios}
\subsubsection{8-dimensional real case}
Now, we allow three of the off-diagonal pairs of entries to be non-zero,
but also require them to be simply real.
We found the separability function
\begin{equation}
\mathcal{S}_{[(1,2),(1,3),(3,4)]}(\nu_{1},\nu_{2}) = 
\mathcal{S}_{[(1,2),(1,4),(3,4)]}(\nu_{1},\nu_{2}) =
\end{equation}
\begin{displaymath}
\begin{cases}
 \frac{4 \pi }{3} & \frac{1}{\nu _1}=\nu _2\land \nu _1>0
   \\
 \frac{4}{3} \pi  \sqrt{\nu _1 \nu _2} & \nu _1>0\land
   \frac{1}{\nu _1}>\nu _2\land \nu _2>0 \\
 \frac{1}{3} \pi  \left(3 \sec ^{-1}\left(\sqrt{\nu _1 \nu
   _2}\right)+4 \sqrt{\nu _1 \nu _2}+\frac{\sqrt{\nu _1
   \nu _2-1}}{\nu _1 \nu _2}-4 \sqrt{\nu _1 \nu
   _2-1}\right) & \nu _1>0\land \frac{1}{\nu _1}<\nu _2
\end{cases}.
\end{displaymath}
Also, we have
\begin{equation}
\mathcal{S}_{[(1,3),(1,4),(2,4)]}(\nu_{1}) =
\begin{cases}
 \frac{4 \pi }{3} & \nu _1=1 \\
 \frac{4 \pi  \sqrt{\nu _1}}{3} & 0<\nu _1<1 \\
 \frac{\pi  \left(4 \nu _1^{3/2}+\left(3 \sin
   ^{-1}\left(\sqrt{1-\frac{1}{\nu _1}}\right)-4 \sqrt{\nu
   _1-1}\right) \nu _1+\sqrt{\nu _1-1}\right)}{3 \nu _1} &
   \nu _1>1
\end{cases}.
\end{equation}
\section{Qutrit-Qutrit Analyses} \label{qutritqutrit}
In the qubit-qubit ($4 \times 4$ density matrix) case, we were able
to express the condition (\ref{Thirdcontion}) 
that the determinant of the partial transpose of
$\rho$ be nonnegative in terms of {\it one} supplementary variable ($\nu$), 
given by (\ref{BlooreRatio}),
rather than three independent diagonal entries.
Similarly, in the qubit-qutrit ($6 \times 6$ density matrix) case,
we could employ {\it two} supplementary variables ($\nu_{1}, \nu_{2}$), 
given by (\ref{tworatios}), rather 
than five independent diagonal entries.

For the qutrit-qutrit ($9 \times 9$ density matrix) case, rather than
eight independent diagonal entries, we found that one can employ
the {\it four} supplementary variables,
\begin{equation}
\nu_{1}= \frac{\rho_{11} \rho_{55}}{\rho_{22} \rho_{44}}; \hspace{.2in}
\nu_{2}= \frac{\rho_{22} \rho_{66}}{\rho_{33} \rho_{55}}; \hspace{.2in}
\nu_{3}= \frac{\rho_{44} \rho_{88}}{\rho_{55} \rho_{77}}; \hspace{.2in}
\nu_{4}= \frac{\rho_{55} \rho_{99}}{\rho_{66} \rho_{88}}.
\end{equation}
\subsection{Thirty-five nullified pairs of off-diagonal entries --- 36 
scenarios}
\subsubsection{10-dimensional complex case --- $P^{HS}_{PPT} =\frac{1}{3}; \frac{1}{6}$}
Here, we nullify all but one of the thirty-six pairs of off-diagonal entries
of the $9 \times 9$ density matrix $\rho$. We allow this solitary pair to be
composed of complex conjugates. Since the Peres-Horodecki 
positive partial 
transposition (PPT) criterion is not sufficient to ensure separability,
we accordingly modify our notation.

Our first result is
\begin{equation}
\mathcal{S}_{[\tilde{(1,5)}]}^{9 \times 9}(\nu_{1}) = 
\begin{cases}
 \pi  & \nu _1\leq 1 \\
 \frac{\pi }{\nu _1} & \nu_{1} > 1
\end{cases}
\end{equation}
(a dual scenario being $[\tilde{(2,4)}]$).
We have $V^{HS}_{tot} =\frac{\pi}{3628800}, V^{HS}_{PPT}= 
\frac{\pi}{10886400}$, so $P^{HS}_{PPT}= \frac{1}{3}$.

The same three outcomes are obtained based on the
PPT function
\begin{equation}
\mathcal{S}_{[\tilde{(1,6)}]}^{9 \times 9}(\nu_{1},\nu_{2}) =
\begin{cases}
 \pi  & \nu_{1} \nu _2 \leq 1 \\
 \frac{\pi }{\nu _1 \nu _2} & \nu _1 \nu_{2} > 1 
\end{cases}.
\end{equation}
On the other hand, we have $V^{HS}_{tot} =\frac{\pi}{3628800}, V^{HS}_{PPT}=
\frac{\pi}{21772800}$, and  $P^{HS}_{PPT}= \frac{1}{6}$ based on the
PPT function
\begin{equation}
\mathcal{S}_{[\tilde{(6,8)}]}^{9 \times 9}(\nu_{4}) =
\begin{cases}
 \pi  & \nu _4\geq 1  \\
 \pi  \nu _4 & 0<\nu _4<1
\end{cases}.
\end{equation}

Of the thirty-six combinatorially possible scenarios, thirteen had
$P^{HS}_{PPT} = \frac{1}{3}$, while four had $P^{HS}_{PPT} = \frac{1}{3}$, 
and the remaining nineteen were fully separable in nature.
\subsection{Thirty-four  nullified pairs of off-diagonal entries --- 630 scenarios}
\subsubsection{12-dimensional complex case --- $P^{HS}_{PPT} =\frac{1}{3}; \frac{7}{30}$}
Since the number of combinatorially possible scenarios was so large,
we randomly generated scenarios to examine.

Firstly, we found
\begin{equation}
\mathcal{S}_{[\tilde{(1,4)},\tilde{(3,5)}]}^{9 \times 9}(\nu_{2}) =
\begin{cases}
 \pi ^2 & \nu _2\geq 1 \\
 \pi ^2 \nu _2 & 0<\nu _2<1
\end{cases}.
\end{equation}
For this scenario, we had 
$V^{HS}_{tot}= \frac{\pi ^2}{479001600}, 
V^{HS}_{PPT} = \frac{\pi^2}{1437004800}$, 
giving us $P^{HS}_{PPT}= \frac{1}{3}$.

Also, we found
\begin{equation}
\mathcal{S}_{[\tilde{(2,9)},\tilde{(6,9)}]}^{9 \times 9}(\nu_{2},\nu_{4}) =
\begin{cases}
 \frac{\pi ^2}{2} & \nu _2  \nu_{4} \leq 1  \\
 \frac{\pi ^2 \left(2 \nu _2 \nu _4-1\right)}{2 \nu _2^2
   \nu _4^2} & \nu _2  \nu_{4} >1
\end{cases}.
\end{equation}
For this scenario, we had
$V^{HS}_{tot}= \frac{\pi ^2}{479001600},
V^{HS}_{PPT} = \frac{\pi^2}{2052864000}$,
giving us $P^{HS}_{PPT}= \frac{7}{30} \approx 0.23333$.
\section{Qubit-Qubit-Qubit Analyses, I} \label{qubitqubitqubitI}
For initial relative simplicity, let us 
regard an $8 \times 8$ density matrix $\rho$ as a {\it bipartite}
system, a composite of a four-level system and a two-level system.
Then, we can compute the partial transposition of $\rho$, transposing in
place its four $4 \times 4$ blocks. 
The nonnegativity of this partial transpose can be expressed using just
{\it three} ratio variables,
\begin{equation}
\nu_{1} = \frac{\rho_{11} \rho_{66}}{\rho_{22} \rho_{55}}; \hspace{.2in}
\nu_{2} = \frac{\rho_{22} \rho_{77}}{\rho_{33} \rho_{66}}; \hspace{.2in}
\nu_{3} = \frac{\rho_{33} \rho_{88}}{\rho_{44} \rho_{77}},
\end{equation}
rather than seven independent diagonal entries.
\subsection{Twenty-seven nullified pairs of off-diagonal entries --- 28
scenarios}
\subsubsection{9-dimensional complex case --- $P^{HS}_{PPT}= \frac{1}{3}$}
We have the PPT function
\begin{equation}
\mathcal{S}_{[\tilde{(1,6)}]}^{8 \times 8}(\nu_{1}) =
\begin{cases}
 \pi  & \nu _1\leq 1 \\
 \frac{\pi }{\nu _1} & \nu_{1} > 1
\end{cases}.
\end{equation}
(Scenario $[\tilde{(2,5)}]$ was dual to this one.)
For this scenario, $V^{HS}_{tot}= \frac{\pi}{362880},
V^{HS}_{PPT}= \frac{\pi}{1088640}$, yielding
$P^{HS}_{PPT}= \frac{1}{3}$.
There were twelve scenarios, {\it in toto}, with 
precisely these three outcomes.
The other sixteen were all fully separable in nature.
\subsection{Twenty-six nullified pairs of off-diagonal entries --- 378
scenarios}
\subsubsection{11-dimensional complex case --- $P^{HS}_{PPT} =\frac{1}{3}; \frac{1}{9}$}
Again, because of the large number of possible scenarios,
we chose them randomly for inspection.

Firstly, we obtained
\begin{equation}
\mathcal{S}_{[\tilde{(3,5)},\tilde{(6,8)}]}^{8 \times 8}(\nu_{1},\nu_{2}) =
\begin{cases}
 \pi ^2 & \nu _1>0\land \frac{1}{\nu _1}\leq \nu _2 \\
 \pi ^2 \nu _1 \nu _2 & \nu _1>0\land \frac{1}{\nu _1}>\nu
   _2\land \nu _2>0
\end{cases}.
\end{equation}
(Of course, the symbols ``$\land$'' and ``$\lor$'', used by Mathematica
in its output, denote the logical connectives ``and'' (conjunction) and ``or''
 (intersection) of propositions.)
For this scenario, we had $V^{HS}_{tot} =
\frac{\pi^2}{39916800}, V^{HS}_{PPT} = \frac{\pi^2}{119750400}$, 
giving us $P^{HS}_{PPT}=\frac{1}{3}$.

Also,
\begin{equation}
\mathcal{S}_{[\tilde{(2,5)},\tilde{(4,7)}]}^{8 \times 8}(\nu_{1},\nu_{2}) =
\begin{cases}
 \pi ^2 & \nu _1\geq 1\land \nu _3\geq 1 \\
 \pi ^2 \nu _1 & 0<\nu _1<1\land \nu _3\geq 1 \\
 \pi ^2 \nu _3 & \nu _1\geq 1\land 0<\nu _3<1 \\
 \pi ^2 \nu _1 \nu _3 & 0<\nu _1<1\land 0<\nu _3<1
\end{cases}.
\end{equation}
For this scenario, we had $V^{HS}_{tot} =
\frac{\pi^2}{39916800}, V^{HS}_{PPT} = \frac{\pi^2}{359251200}$,
giving us $P^{HS}_{PPT}=\frac{1}{9}$.

We also found the PPT function
\begin{equation}
\mathcal{S}_{[\tilde{(1,3)},\tilde{(4,7)}]}^{8 \times 8}(\nu_{1}) =
\begin{cases}
 \frac{\pi ^2}{2} & \nu _3=1 \\
 \frac{\pi ^2 \nu _3}{2} & 0<\nu _3<1 \\
 \pi  \left(\cos ^{-1}\left(\sqrt{1-\frac{1}{\nu
   _3}}\right)-\csc ^{-1}\left(\sqrt{\nu _3}\right)\right)
   \nu _3+\pi ^2-\frac{\pi ^2}{2 \nu _3} & \nu _3>1
\end{cases}.
\end{equation}

\section{Qubit-Qubit-Qubit Analyses. II} \label{qubitqubitqubitII}
Here we regard the $8 \times 8$ density matrix as a {\it tripartite}
composite of three two-level systems, and compute the partial transpose by
transposing in place the eight $2 \times 2$ blocks of $\rho$.
(For {\it symmetric} states of three qubits, positivity of the partial
transpose is {\it sufficient} to ensure separability \cite{eckert}.)
Again the nonnegativity of the determinant could be expressed using
three (different) ratio variables,
\begin{equation}
\nu_{1} = \frac{\rho_{11} \rho_{44}}{\rho_{22} \rho_{33}}; \hspace{.2in}
\nu_{2} = \frac{\rho_{44} \rho_{55}}{\rho_{33} \rho_{66}}; \hspace{.2in}
\nu_{3} = \frac{\rho_{55} \rho_{88}}{\rho_{66} \rho_{77}},
\end{equation}
\subsection{Twenty-seven nullified pairs of off-diagonal entries --- 28
scenarios}
\subsubsection{9-dimensional complex case --- $P^{HS}_{PPT}= \frac{1}{3}$}
There were, again, twelve of twenty-eight scenarios with non-trivial
separability properties, all with 
 $V^{HS}_{tot}= \frac{\pi}{362880},
V^{HS}_{PPT}= \frac{\pi}{1088640}$, yielding
$P^{HS}_{PPT}= \frac{1}{3}$. 
One of these was
\begin{equation}
\mathcal{S}_{[\tilde{(1,4)}]}^{8 \times 8}(\nu_{1}) =
\begin{cases}
 \pi  & \nu _1\leq 1 \\
 \frac{\pi }{\nu _1} & \nu_{1} > 1
\end{cases}.
\end{equation}
\subsubsection{11-dimensional complex case --- $P^{HS}_{PPT}= \frac{17}{60}; \frac{1}{3}$}
We obtained the PPT function
\begin{equation}
\mathcal{S}_{[\tilde{(1,8)},\tilde{(5,7)}]}^{8 \times 8}(\nu_{1},\nu_{2},\nu_{3}) =
\begin{cases}
 \frac{\pi ^2}{2} & \frac{\nu _2}{\nu _1}=\nu _3\land \nu
   _1>0\land \nu _2>0 \\
 \pi ^2 & \nu _2>0\land \left(\left(\nu _1=0\land \nu
   _3\geq 0\right)\lor \left(\nu _3=0\land \nu
   _1>0\right)\right) \\
 \frac{\pi ^2 \nu _2}{4 \nu _1 \nu _3} & \nu _1>0\land \nu
   _2>0\land \frac{\nu _2}{\nu _1}<\nu _3 \\
 \pi ^2-\frac{\pi ^2 \nu _1 \nu _3}{2 \nu _2} & \nu
   _1>0\land \nu _2>0\land \frac{\nu _2}{\nu _1}>\nu
   _3\land \nu _3>0
\end{cases}.
\end{equation}
For this we had $V_{tot}^{HS}= \frac{\pi^2}{39916800},
V_{PPT}^{HS} =\frac{17 \pi^2}{239500800}$, giving us
$P^{HS}_{PPT}= \frac{17}{60} \approx 0.283333$.

Additionally,
\begin{equation}
\mathcal{S}_{[\tilde{(1,4)},\tilde{(7,8)}]}^{8 \times 8}(\nu_{1})= 
\begin{cases}
 \pi ^2 & \nu _1\leq 1 \\
 \frac{\pi ^2}{\nu _1} & \nu_{1} > 1.
\end{cases}
\end{equation}
Here, we had $V_{tot}^{HS}= \frac{\pi^2}{39916800},
V_{PPT}^{HS} =\frac{17 \pi^2}{119750400}$, giving us
$P^{HS}_{PPT}= \frac{1}{3}$.

Another PPT function we were able to find was
\begin{equation}
\mathcal{S}_{[\tilde{(3,4)},\tilde{(3,8)}]}^{8 \times 8}(\nu_{2},\nu_{3})=
\begin{cases}
 \frac{\pi ^2}{2} & \nu _2>0\land \left(\nu _3=\nu _2\lor
   \left(\nu _2>\nu _3\land \nu _2<2 \nu _3\right)\lor
   \left(\nu _2\geq 2 \nu _3\land \nu _3\geq
   0\right)\right) \\
 \frac{\pi ^2 \nu _2 \left(2 \nu _3-\nu _2\right)}{2 \nu
   _3^2} & \nu _2>0\land \nu _2<\nu _3
\end{cases}.
\end{equation}

\section{Approximate Approaches to 9-Dimensional Real Qubit-Qubit Scenario} \label{approximate}
\begin{figure}
\includegraphics{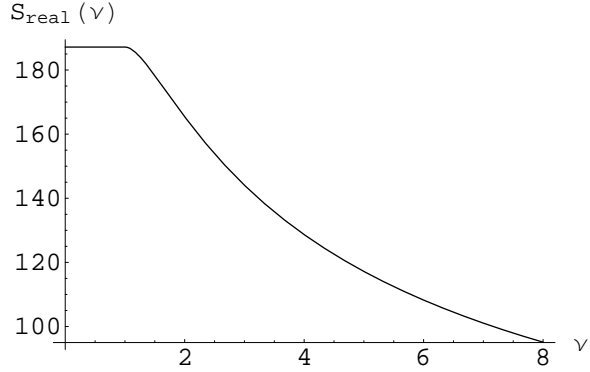}
\caption{\label{fig:firstapprox}Approximation (\ref{firstminorapprox}) to the separability
function for the 9-dimensional real two-qubit states}
\end{figure}

As we have earlier emphasized (sec.~\ref{Strz}), it appears that the 
simultaneous computational 
enforcement of the four conditions 
((\ref{firstcondition}), (\ref{secondcondition}), 
(\ref{2by2}), (\ref{Thirdcontion})) 
that would yield us the 9-dimensional
volume of the separable real two-qubit states appears presently 
highly intractable.
But if we replace (\ref{Thirdcontion})  by {\it less} strong conditions on the
nonnegativity of the partial transpose ($\rho^{T}$), 
we can achieve some form of
approximation to the desired results. So, replacing 
(\ref{Thirdcontion}) 
by the requirement (derived from a $2 \times 2$ principal minor 
of $\rho^{T}$) that 
\begin{equation} \label{minor1}
1 - \nu z_{14}^2 \geq 0,
\end{equation} 
we obtain the
{\it approximate} separability function (Fig.~\ref{fig:firstapprox})
\begin{equation} \label{firstminorapprox}
S_{real}(\nu)=
\begin{cases}
 \frac{512 \pi ^2}{27} & 0 <  \nu \leq 1 \\
 \frac{256 \left(3 \pi ^2 \nu -\pi ^2\right)}{27 \nu
   ^{3/2}} & \nu > 1
\end{cases}.
\end{equation}
(In the analyses in this section, we utilize the integration limits
on the $z_{ij}$'s \cite[eqs. (3)-(5)]{slaterPRA2} 
yielded by the cylinrical decomposition algorithm [CAD], to reduce the
dimensionalities of our constrained integrations.)
This yields an {\it upper bound} on the separability probability
of the real 9-dimensional qubit-qubit states of
$\frac{1}{2}+\frac{512}{135 \pi ^2} \approx 0.88427$. 
We obtain the same probability if we employ instead of (\ref{minor1}) the
requirement
\begin{equation} \label{minor2}
\nu - z_{23}^2 \geq 0,
\end{equation}
which yields the dual function to (\ref{firstminorapprox}), namely,
\begin{equation} \label{secondminorapprox}
S_{real}(\nu)=
\begin{cases}
 \frac{512 \pi ^2}{27} & \nu \geq 1 \\
 -\frac{256}{27} \pi ^2 (\nu -3) \sqrt{\nu } & 0<\nu <1
\end{cases}.
\end{equation}
(The left-hand sides of (\ref{minor1}) and (\ref{minor2}) are the only
two of the six $2 \times 2$ principal minors of $\rho^{T}$ that are
non-trivially distinct --- apart from cancellable 
nonnegative factors --- from the corresponding minors of $\rho$ itself.)
If we form a ``quasi-separability'' function by piecing together
the non-constant segments of (\ref{minor1}) and (\ref{minor2}), we can
infer --- using a simple symmetry, duality argument --- an improved lower bound on the HS separability probability of
$\frac{1024}{135 \pi^2} \approx 0.76854$.
\subsection{Analytically-derived beta functions}
We have not, to this point of time, been able to {\it jointly} enforce
the constraints (\ref{minor1}) and (\ref{minor2}). (Doing so, would amount to 
solving a constrained {\it four}-dimensional integration problem.) However, we had some
interesting success in 
jointly enforcing  (\ref{minor1}) along with the constraint
\begin{equation}
-z_{12}^2+2 \sqrt{\nu } z_{13} z_{14} z_{12}-z_{13}^2-\nu 
   z_{14}^2+1 \geq 0,
\end{equation}
corresponding to the first of the four principal $3 \times 3$ 
minors  of $\rho^{T}$. (This amounts to solving a constrained {\it three}-dimensional integration problem.)
In attempting to jointly enforce these two constraints, we obtained
(without, however, being {\it directly} 
able to explicitly derive 
the separability function 
over $\mu \in [0,\infty]$) an upper bound on $V^{HS}_{sep}$ (and a 
consequent upper bound on $P^{HS}_{sep}$ 
of 0.77213) of
\begin{equation}
\frac{11 \pi ^4}{2116800} + 
\int_{0}^{1} 
\hat{\mathcal{J}}_{real}(\mu) s_{real}(\mu) d \mu \approx  0.00124359.
\end{equation}
Here we have the approximate separability function 
(monotonically-{\it decreasing} for $\mu \in [0,1]$),
\begin{equation} \label{quasi}
s_{real}(\mu) =\frac{\pi ^3 \left(13 B_{\mu
   ^2}\left(\frac{1}{2},\frac{3}{2}\right) \mu ^2+\left(2
   \mu ^2-13\right) B_{\mu
   ^2}\left(\frac{3}{2},\frac{3}{2}\right)-2 B_{\mu
   ^2}\left(\frac{5}{2},\frac{3}{2}\right)\right)}{3 \mu
   ^3}.
\end{equation}
For (apparently necessary) 
convenience in computation, so that Mathematica only has to deal with 
integral powers rather than 
half-integer ones, we use $\mu = \sqrt{\nu}$ (as we did in our
original study \cite{univariate})
and the corresponding jacobian function \cite[eq. (10)]{univariate} 
(cf. (\ref{Jacreal})) is
\begin{equation}
\hat{\mathcal{J}}_{real}(\mu) =
\frac{\mu ^4 \left(12 \left(\left(\mu ^2+2\right)
   \left(\mu ^4+14 \mu ^2+8\right) \mu ^2+1\right) \log
   (\mu )-5 \left(5 \mu ^8+32 \mu ^6-32 \mu
   ^2-5\right)\right)}{1890 \left(\mu ^2-1\right)^9}.
\end{equation}
What is most interesting here, of course, 
is that we have 
beta functions (now derived {\it analytically}) 
explicitly appearing in the (approximate) separability 
function formula (\ref{quasi}), while we only previously had
{\it numerical} evidence for their relevance \cite{slaterPRA2}.

Though we sought to enforce additional constraints based on the principal
minors of $\rho^{T}$, in order to obtain tighter upper bounds on
$V^{HS}_{sep}$ and 
$P^{HS}_{sep}$, this did not seem to be computationally possible. 
(One might also pursue similar ``approximate'' strategies in the lower-dimensional instances, studied earlier in this paper, still not apparently 
amenable to exact solutions.)

Also, as was mentioned in the Introduction, we have not yet been able to 
exploit to derive HS separability functions and probabilities, the simplified integration limits (\ref{newlimits}) based on the spheroidal
parameterization (\ref{reparameterization}) 
suggested by the geometric discussion of Bloore \cite[secs. 6, 7]{bloore}.
\section{Concluding Remarks} \label{concludingremarks}
Though we have regarded the derivation of the scenario-specific
separability (and PPT) 
{\it functions} as our primary theoretical and computational
challenge, the derivation of the HS separability (and PPT) 
{\it probabilities} 
from such functions --- that is, the performance of an 
integration over the (n-1)-dimensional simplex spanned by the 
diagonal entries of the $n \times n$ density matrix $\rho$ --- can 
also be quite difficult.
(As a general, somewhat informal 
observation, real off-diagonal entries in our scenarios
tend to yield square roots of the diagonal entries in the integrands, 
which prove more challenging to integrate than complex off-diagonal entries.)

The qubit-qubit results above motivated us to reexamine 
previously obtained results
(cf. \cite[eqs. (12), (13)]{univariate}) and we would like to 
make the following observations pertaining to
the {\it full} 9-dimensional real and 15-dimensional HS separability 
probability issue.
We have the exact results in these two cases that
\begin{equation} \label{jacrealInt}
\int_{0}^{\infty} \mathcal{J}_{real}(\nu) = 2 \int_{0}^{1} \mathcal{J}_{real}(\nu) =\frac{\pi^2}{1146880} \approx 8.60561 \cdot 10^{-6}
\end{equation}
and
\begin{equation} \label{jaccomplexInt}
\int_{0}^{\infty} \mathcal{J}_{complex}(\nu) = 2 \int_{0}^{1} \mathcal{J}_{complex}(\nu) =\frac{1}{1009008000} \approx 
9.91072 \cdot 10^{-10} .
\end{equation}
Now, to obtain the corresponding {\it total} (separable plus nonseparable) 
 HS volumes \cite{szHS}, that is,
$\frac{\pi^4}{60480} \approx 0.0016106$ 
and $\frac{\pi^6}{851350500} \approx 1.12925 \cdot 10^{-6}$, one must multiply
(\ref{jacrealInt}) and (\ref{jaccomplexInt}) by the 
factors of $C_{real}= \frac{512 \pi^2}{27} = \frac{2^8 \pi^2}{3^3} \approx  
187.157$ and 
$C_{complex} 
\frac{32 \pi^6}{27} = \frac{2^5 \pi^6}{3^3} \approx 1139.42$, respectively.

To most effectively compare these previously-reported results with 
those derived above in this paper, one needs to multiply 
 $C_{real}$ and $\mathcal{S}_{real}(\nu)$,
by $2^{-4}=\frac{1}{16}$ and in the complex case by
$2^{-7}=\frac{1}{128}$. Doing so, for example, 
would adjust $C_{real}$ to equal
$c_{real} = \frac{32 \pi^2}{27} \approx 11.6973$, which we note, in line with
our previous series of calculations [sec.~\ref{sec8dim}] 
is equal to $\frac{4}{3} (\frac{8 \pi^2}{9}$). (Andai \cite{andai} 
also computed the same
volumes --- up to a normalization factor --- as 
Sommers and \.Zyczkowski \cite{szHS}.)
Now, our estimates from \cite{slaterPRA2} are that 
$\mathcal{S}_{real}(1)= 114.62351 < C_{real}$ and
$\mathcal{S}_{complex}(1) = 387.50809 < C_{complex}$. 
These results would appear --- as remarked above --- to 
be a reflection of the phenomena that there
are non-separable states for both the 9- and 15-dimensional scenarios
at $\nu=1$ (the locus of the fully mixed, classical state).

If we take as a trial function,
\begin{equation}
 \mathcal{S}_{real}(\nu) = 114.62351 \left(\begin{cases}
 \sqrt{\nu} & 0\leq \nu\leq 1 \\
 \frac{1}{\sqrt{\nu}} & \nu>1
\end{cases}
\right)
\end{equation}
and insert it into our formula
(\ref{Vreal}), as at least the 
lower-dimensional real results in Fig.~\ref{fig:good}  
and (\ref{suggestion}) might suggest
trying, we obtain the outcome, $V^{HS}_{sep/real} = 0.000613694$
and (using the {\it known} $V^{HS}_{tot/real}$ \cite{szHS}) 
an associated HS separability probability of  
0.381034. This --- as we expected --- is 
certainly not in accord with the {\it numerical} results
of \cite[sec.~V.A.2]{slaterPRA2}. 
The HS real separability probability was estimated there to be
$P^{HS}_{sep/real} = 0.4538838$ and $\mathcal{S}_{real}(\nu)$ (cf. (\ref{Freal})) was 
estimated 
to be closely proportional  to the incomplete beta function 
$B_{\nu}(\frac{1}{2},\sqrt{3})$ and {\it not} the quite different function 
$\sqrt{\nu} \propto B_{v}({\frac{1}{2}},1)$, as our small exercise 
here assumes.

So, we may say, in partial summary that we have been able to obtain 
certain {\it exact}
two-qubit HS separability probabilities in dimensions 
seven or less, making use of the 
advantageous Bloore parameterization \cite{bloore}, but not yet in
dimensions greater than seven. This, however, is considerably greater than 
simply the three dimensions (parameters)
we were able to achieve \cite{pbsJak} in a somewhat comparable study based on 
the {\it generalized Bloch representation} parameterization \cite{kk,jak}.
In \cite{pbsJak} --- extending an approach of Jak\'obczyk and 
Siennicki \cite{jak} --- we 
primarily studied {\it two}-dimensional sections of a set of 
generalized Bloch vectors corresponding to $n \times n$ density matrices, 
for $n=4, 6, 8, 9$ and 10. For $n>4$, by far the most frequently recorded
HS separability [or positive partial transpose (PPT) for $n>6$]
probability was $\frac{\pi}{4} \approx 0.785398$. A very wide range
of exact HS separability and PPT probabilities were tabulated.

Immediately below is just one of many matrix tables 
(this one being numbered (5) in \cite{pbsJak}) presented in \cite{pbsJak} 
(which due to its large size has been left 
simply as a preprint, rather than 
submitted directly to a journal).
This table gives the HS separability probabilities for the
qubit-qutrit case. In the first column are given the identifying numbers
of a {\it pair} of generalized Gell-mann matrices (generators of $SU(6)$).
In the second column of (\ref{n=6case1}) 
are shown the {\it number} of distinct unordered pairs
of $SU(6)$ generators which share the same total (separable and nonseparable)
HS volume, as well as the same separable HS volume, and consequently,
identical HS separability probabilities. The third column gives us these
HS total volumes, the fourth column, the HS separability probabilities and 
the last (fifth) column, numerical approximations to the exact probabilities 
(which, of course, we see --- being probabilities --- do not exceed the  value 1).
 (Due to space/page width constraints, we were unable to generally present
in these data arrays 
the HS separable volumes too, 
though they can, of course, be deduced from the total volume and the separability probability.)

\begin{equation} \label{n=6case1}
\left(
\begin{array}{lllll}
 \{1,13\} & 48 & \frac{4}{9} & \frac{\pi }{4} & 0.785398
   \\
 \{3,11\} & 4 & \frac{8 \sqrt{2}}{27} & \frac{1}{\sqrt{2}}
   & 0.707107 \\
 \{3,13\} & 4 & \frac{4}{9} & \frac{5}{6} & 0.833333 \\
 \{3,25\} & 4 & \frac{8 \sqrt{2}}{27} & \frac{5}{4
   \sqrt{2}} & 0.883883 \\
 \{8,13\} & 4 & \frac{2}{3} & \frac{1}{\sqrt{3}} &
   0.577350 \\
 \{8,25\} & 4 & \frac{\sqrt{2}}{3} & \sqrt{\frac{2}{3}} &
   0.816497 \\
 \{11,15\} & 4 & \frac{4 \sqrt{2} \pi }{27} &
   \frac{1}{3}+\frac{3 \sqrt{3}}{4 \pi } & 0.746830 \\
 \{11,24\} & 2 & \frac{25 \sqrt{\frac{5}{2}}}{72} &
   \frac{2}{5}+\frac{1}{2} \sin
   ^{-1}\left(\frac{4}{5}\right) & 0.863648 \\
 \{13,24\} & 2 & \frac{25 \sqrt{\frac{5}{2}}}{72} &
   \frac{8}{75} \left(-2+5 \sqrt{5}\right) & 0.979236 \\
 \{13,35\} & 4 & \frac{4 \sqrt{\frac{3}{5}}}{5} &
   \frac{1}{12} \left(5+3 \sqrt{5} \csc
   ^{-1}\left(\frac{3}{\sqrt{5}}\right)\right) & 0.886838
   \\
 \{15,16\} & 4 & \frac{32 \sqrt{2}}{81} & \frac{1}{32}
   \left(9 \sqrt{3}+4 \pi \right) & 0.879838 \\
 \{16,24\} & 2 & \frac{25}{144} \sqrt{\frac{5}{2}} \pi  &
   \frac{4+5 \sin ^{-1}\left(\frac{4}{5}\right)}{5 \pi } &
   0.549815 \\
 \{20,24\} & 2 & \frac{25}{144} \sqrt{\frac{5}{2}} \pi  &
   \frac{92+75 \sin ^{-1}\left(\frac{4}{5}\right)}{75 \pi
   } & 0.685627 \\
 \{24,25\} & 2 & \frac{25}{27 \sqrt{2}} & 1-\frac{2}{5
   \sqrt{5}} & 0.821115 \\
 \{24,27\} & 2 & \frac{25}{27 \sqrt{2}} & \frac{92+75 \cos
   ^{-1}\left(\frac{3}{5}\right)}{80 \sqrt{5}} & 0.903076
   \\
 \{25,35\} & 4 & \frac{\sqrt{3} \pi }{5} &
   \frac{\sqrt{5}+3 \csc
   ^{-1}\left(\frac{3}{\sqrt{5}}\right)}{3 \pi } &
   0.504975
\end{array}
\right).
\end{equation}
It might be of interest to address separability problems that appear to be
computationally intractable in the generalized Bloch representation by
transforming them into the Bloore parameterization.

We leave the reader with the intriguing questions, still left unanswered:
do simple, exact formulas exist for 
the Hilbert-Schmidt --- and/or Bures, Kubo-Mori, 
Wigner-Yanase, \ldots (cf. \cite{slaterJGP}) --- separability probabilities for the full 9-dimensional 
and 15-dimensional convex sets of  real and 
complex $4 \times 4$ density matrices 
(and the 20-dimensional and 35-dimensional convex sets of real and 
complex $6 \times 6$ density matrices)? Further, are there helpful
intuitions --- above and beyond the direct implementation 
of our formulas (\ref{Vsmall}) and (\ref{Vbig}) --- that 
can aid in understanding the abundance of elegant,
simple results occurring in this general research area?

\begin{acknowledgments}
I would like to express gratitude to the Kavli Institute for Theoretical
Physics (KITP)
for computational support in this research.

\end{acknowledgments}

\bibliography{Zest1}

\end{document}